\documentstyle{amsppt}
\newcount\mgnf\newcount\tipi\newcount\tipoformule\newcount\greco 
\tipi=2          
\tipoformule=0   

\global\newcount\numsec\global\newcount\numfor
\global\newcount\numapp\global\newcount\numcap
\global\newcount\numfig\global\newcount\numpag
\global\newcount\numnf
\global\newcount\numtheo

\def\SIA #1,#2,#3 {\senondefinito{#1#2}%
\expandafter\xdef\csname #1#2\endcsname{#3}\else
\write16{???? ma #1,#2 e' gia' stato definito !!!!} \fi}

\def \FU(#1)#2{\SIA fu,#1,#2 }

\def\etichetta(#1){(\veroparagrafo.\veraformula)%
\SIA e,#1,(\veroparagrafo.\veraformula) %
\global\advance\numfor by 1%
\write15{\string\FU (#1){\equ(#1)}}%
\write16{ EQ #1 ==> \equ(#1)  }}

\def\etichettat(#1){\veroparagrafo.\veratheorema:%
\SIA e,#1,{\veroparagrafo.\veratheorema} %
\global\advance\numtheo by 1%
\write15{\string\FU (#1){\thu(#1)}}%
\write16{ TH #1 ==> \thu(#1)  }}

\def\etichettaa(#1){(A\veraappendice.\veraformula)
 \SIA e,#1,(A\veraappendice.\veraformula)
 \global\advance\numfor by 1
 \write15{\string\FU (#1){\equ(#1)}}
 \write16{ EQ #1 ==> \equ(#1) }}
\def\getichetta(#1){Fig. \verafigura
 \SIA g,#1,{\verafigura}
 \global\advance\numfig by 1
 \write15{\string\FU (#1){\graf(#1)}}
 \write16{ Fig. #1 ==> \graf(#1) }}
\def\retichetta(#1){\numpag=\pgn\SIA r,#1,{\verapagina}
 \write15{\string\FU (#1){\rif(#1)}}
 \write16{\rif(#1) ha simbolo  #1  }}
\def\etichettan(#1){(n\verocapitolo.\veranformula)
 \SIA e,#1,(n\verocapitolo.\veranformula)
 \global\advance\numnf by 1
\write16{\equ(#1) <= #1  }}

\newdimen\gwidth
\gdef\profonditastruttura{\dp\strutbox}
\def\senondefinito#1{\expandafter\ifx\csname#1\endcsname\relax}
\def\BOZZA{
\def\alato(##1){
 {\vtop to \profonditastruttura{\baselineskip
 \profonditastruttura\vss
 \rlap{\kern-\hsize\kern-1.2truecm{$\scriptstyle##1$}}}}}
\def\galato(##1){ \gwidth=\hsize \divide\gwidth by 2
 {\vtop to \profonditastruttura{\baselineskip
 \profonditastruttura\vss
 \rlap{\kern-\gwidth\kern-1.2truecm{$\scriptstyle##1$}}}}}
\def\verapagina{
{\romannumeral\number\numcap}.\number\numsec.\number\numpag}}

\def\alato(#1){}
\def\galato(#1){}
\def\veroparagrafo{\number\numsec}\def\veraformula{\number\numfor}
\def\veraappendice{\number\numapp}
\def\verapagina{\number\pageno}\def\veranformula{\number\numnf}
\def\verafigura{{\romannumeral\number\numcap}.\number\numfig}
\def\verocapitolo{\number\numcap}\def\veranformula{\number\numnf}
\def\veratheorema{\number\numtheo}
\def\Eqn(#1){\eqno{\etichettan(#1)\alato(#1)}}
\def\eqn(#1){\etichettan(#1)\alato(#1)}
\def\TH(#1){{\etichettat(#1)\alato(#1)}}
\def\thv(#1){\senondefinito{fu#1}$\clubsuit$#1\else\csname fu#1\endcsname\fi} 
\def\thu(#1){\senondefinito{e#1}\thv(#1)\else\csname e#1\endcsname\fi}
\def\ver{\veroparagrafo}
\def\Eq(#1){\eqno{\etichetta(#1)\alato(#1)}}
\def\eq(#1){\etichetta(#1)\alato(#1)}
\def\Eqa(#1){\eqno{\etichettaa(#1)\alato(#1)}}
\def\eqa(#1){\etichettaa(#1)\alato(#1)}
\def\dgraf(#1){\getichetta(#1)\galato(#1)}
\def\drif(#1){\retichetta(#1)}

\def\eqv(#1){\senondefinito{fu#1}$\clubsuit$#1\else\csname fu#1\endcsname\fi}
\def\equ(#1){\senondefinito{e#1}\eqv(#1)\else\csname e#1\endcsname\fi}
\def\graf(#1){\senondefinito{g#1}\eqv(#1)\else\csname g#1\endcsname\fi}
\def\rif(#1){\senondefinito{r#1}\eqv(#1)\else\csname r#1\endcsname\fi}
\def\bib[#1]{[#1]\numpag=\pgn
\write13{\string[#1],\verapagina}}

\def\include#1{
\openin13=#1.aux \ifeof13 \relax \else
\input #1.aux \closein13 \fi}

\openin14=\jobname.aux \ifeof14 \relax \else
\input \jobname.aux \closein14 \fi
\openout15=\jobname.aux
\openout13=\jobname.bib


\ifnum\tipoformule=1\let\Eq=\eqno\def\eq{}\let\Eqa=\eqno\def\eqa{}
\def\equ{}\fi


{\count255=\time\divide\count255 by 60 \xdef\hourmin{\number\count255}
        \multiply\count255 by-60\advance\count255 by\time
   \xdef\hourmin{\hourmin:\ifnum\count255<10 0\fi\the\count255}}

\def\oramin{\hourmin }

\def\data{\number\day/\ifcase\month\or january \or february \or march \or
april \or may \or june \or july \or august \or september
\or october \or november \or december \fi/\number\year;\ \oramin}

\setbox200\hbox{$\scriptscriptstyle \data $}

\newcount\pgn \pgn=1
\def\foglio{\number\numsec:\number\pgn
\global\advance\pgn by 1}
\def\foglioa{A\number\numsec:\number\pgn
\global\advance\pgn by 1}

\footline={\rlap{\hbox{\copy200}}\hss\tenrm\folio\hss}


\global\newcount\numpunt

\magnification=\magstephalf
\baselineskip=16pt
\parskip=8pt

\voffset=2.5truepc
\hoffset=0.5truepc
\hsize=6.1truein
\vsize=8.4truein 
{\headline={\ifodd\pageno\rightheadline \else \leftheadline \fi}}
\def\rightheadline{\it  {Fluctuations}\hfil\tenrm\folio}
\def\leftheadline{\tenrm \folio \hfil\it  {Section $\ver$}}

\def\a{\alpha}
\def\b{\beta}
\def\d{\delta}
\def\e{\epsilon}

\def\f{\phi}

\def\s{\sigma}

\def\1{{1\kern-.25em\roman{I}}}
\def\eu{{1\kern-.25em\roman{I}}}
\def\f1{{1\kern-.25em\roman{I}}}

\def\R{{\Bbb R}}  
\def\N{{\Bbb N}}  
\def\P{{\Bbb P}}  
\def\Z{{\Bbb Z}}  
\def\E{{\Bbb E}}  




\let\cal=\Cal

\def\DD{{\cal D}}

\def\NN{{\cal N}}

\def\PP{{\cal P}}

\def\RR{{\cal R}}
\def\SS{{\cal S}}

\def\VV{{\cal V}}

\def\VV{{\cal V}}
\def\WW{{\cal W}}

\def\YY{{\cal Y}}

\def\chap #1#2{\line{\ch #1\hfill}\numsec=#2\numfor=1\numtheo=1}

\def\wt{\widetilde}

\def\limlaw{\buildrel \DD\over\rightarrow}


\def\note#1{\footnote{#1}}

\def\frac#1#2{{#1\over #2}}

\def\text#1{\quad{\hbox{#1}}\quad}

\def\proposition #1{\noindent{\thbf Proposition #1}}

\def\theo #1{\noindent{\thbf Theorem {#1} }}

\def\lemma #1{\noindent{\thbf Lemma {#1} }}

\def\corollary #1{\noindent{\thbf Corollary #1 }}

\def\endproof{$\diamondsuit$}
\def\remark{\noindent{\bf Remark: }}
\def\thanks{\noindent{\bf Acknowledgements: }}

\font\thbf=cmbxsl10 scaled\magstephalf

\font\ch=cmbx12

\font\it=cmti10
\font\bf=cmbx10


\overfullrule=0pt

\font\tit=cmbx12
\font\aut=cmbx12

\centerline{\tit FLUCTUATIONS OF THE FREE ENERGY IN THE }
\vskip.2truecm
\centerline{\tit REM AND THE $P$-SPIN SK MODELS}
\vskip1truecm 

\centerline{\aut Anton Bovier 
\note{ Weierstrass-Institut f\"ur Angewandte Analysis und Stochastik,
Mohrenstrasse 39, D-10117 Berlin,\hfill\break Germany.
 e-mail: bovier\@wias-berlin.de},
Irina Kurkova\note{EURANDOM, P.O. Box 513 - 5600 MB Eindhoven, The Netherlands
.\hfill\break
e-mail: kurkova\@eurandom.tue.nl},}
\centerline{\aut Matthias L\"owe\note{
EURANDOM, P.O. Box 513 - 5600 MB Eindhoven, The Netherlands.\hfill\break
email: Lowe\@eurandom.tue.nl}}
\vskip0.5cm

\vskip0.5truecm\rm

\noindent {\bf Abstract:} We consider the random fluctuations of the
free energy in the $p$-spin version of the Sherrington-Kirkpatrick 
model in the high temperature regime. Using the martingale approach 
of Comets and Neveu as used in the standard SK model combined with 
truncation techniques inspired by a recent paper by Talagrand on the 
$p$-spin version, we prove that (for $p$ even)
 the random corrections to the 
free energy are on  a scale $N^{-(p-2)/4}$ only, and after proper 
rescaling converge to a standard Gaussian random variable. This is
shown to hold for all values of the inverse temperature, $\b$, smaller than 
a critical $\b_p$. We also show that $\b_p\rightarrow \sqrt{2\ln 2}$ as
$p\uparrow +\infty$.  Additionally we study the
formal $p\uparrow +\infty$ limit of these models, the random energy model. 
Here we compute the precise limit theorem for the partition 
function at {\it all}
temperatures. For $\b<\sqrt{2\ln2}$, fluctuations are found
at an {\it exponentially small} scale, with two distinct limit laws
above and below a second critical value $\sqrt{\ln 2/2}$: For $\b$ up to that
  value the rescaled fluctuations are Gaussian, while below that there are 
non-Gaussian fluctuations driven by the Poisson process of 
the extreme values of the random energies.
For $\b$ larger than the critical $\sqrt{2\ln 2}$, the fluctuations of the
logarithm of the partition function are on scale one and are expressed in terms
of the Poisson process of extremes. At the critical temperature,
the partition function divided by its expectation converges to $1/2$.


 \vfill
$ {} $


\chap{1. Introduction.}1

In recent years it has become increasingly clear that a problem
of central importance for the understanding of disordered spin systems 
is the control of random fluctuations of thermodynamic quantities
[AW,NS,BM,T1]. Unfortunately, a precise control of such quantities is very 
hard to come by. Concentration of measure techniques [T2] have been 
realized to be efficient tools to get {\it upper bounds}
[BGP1,BG1], but lower bounds or exact limit theorems are scarce.
One of these examples is the Sherrington-Kirkpatrick (SK) model in the 
high-temperature phase, where a central limit theorem for the free energy
was proven first by Aizenman, Lebowitz and Ruelle [ALR], using 
cluster expansion techniques, and later by Comets and Neveu [CN], making use of
martingale methods and stochastic calculus. Their methods have been extended
to a few related cases [Tou,B1] later. In the present paper we want to 
continue this effort by investigating a large class of natural generalisation
of the SK model, the so called $p$-spin SK models, and their $p\uparrow +\infty$
limit, the random energy model (REM).

For our present purposes it is natural to consider the class of models
we study as Gaussian processes on the hypercube $\SS_N=\{-1,1\}^N$. We will
always denote the corner of $\SS_N$ by $\s$; for historical reasons they
 are called {\it spin configurations}. A Gaussian process $X$ on $\SS_N$
is characterized completely by its mean and covariance function. 
The processes
we consider will always be assumed to have mean zero and covariance
$$
\E X_\s X_{\s'} \equiv f (R_N(\s,\s')),
\Eq(1.1)
$$
where $f$ depends on the so-called {\it overlap}\note{The overlap 
is related to the Hamming distance $d_{Ham}$ by $d_{Ham}(\s,\s') =
N(1-R_N(\s,\s'))/2$.},
$R_N(\s,\s')\equiv N^{-1}(\s,\s')\equiv N^{-1}\sum_{i=1}^N \s_i\s'{}_i$,
In this note we will concentrate on the case where
$f(x)= f_p(x):=x^{p}$, with $p$ even.\note{The case $p$ odd can also be treated, but 
presents considerable additional computational problems.}
In this case, $X_\s$ can be represented in the form 
$$
X_\s= N^{-p/2}\sum_{i_1,i_2,\ldots,i_p} J_{i_1,i_2,\dots,i_p} \s_{i_1}
\s_{i_2}\cdots
\s_{i_p}
\Eq(1.2)
$$
with $J_{i_1,\dots,i_p}$ a family of i.i.d.\ normal random variables. Since for
$p=2$ we obtain the classical SK model, this 
representation justifies the name {\it $p$-spin SK model}. 
Note that as $p$ increases, the process gets more and more de-correlated,
and in the limit $p\uparrow +\infty$ we arrive at the case where $X_\s$ are
i.i.d.\ normal random variables.

Given such a Gaussian process, our main object of interest is the 
so called {\it partition function},
$$
Z_{\b,N}\equiv \E_\s e^{\b \sqrt N X_\s}\equiv 2^{-N}\sum_{\s\in\SS_N}
 e^{\b \sqrt N X_\s}.
\Eq(1.3)
$$
The quantities
$ e^{\b \sqrt N X_\s}$ are called {\it Boltzmann weights } and the 
parameter $\b\in \R_+$ 
is known as the {\it inverse temperature}, and 
$H_N(\s)\equiv\sqrt N X_\s$  as (minus) the 
{\it Hamiltonian} in statistical 
mechanics. $Z_{\b,N}$  are random variables, and we are primarily interested
in their behaviour as $N$ tends to infinity. In statistical mechanics, it 
is customary to introduce the so-called {\it free energy}
$$
F_{\b,N}\equiv -\frac 1{\b N} \ln Z_{\b,N}.
\Eq(1.4)
$$
It is easy to prove in all the models we consider here, that for all values 
of $\b$, $F_{\b,N}$ is a {\it self-averaging} quantity, i.e.\ that
$$
\lim_{N\uparrow + \infty} \left|F_{\b,N}-\E F_{\b,N}\right| =0
\text{a.s.}
\Eq(1.5)
$$
It is, however, not known in general whether the so called {\it quenched}
free energy $\E F_{\b,N} $ converges to a limit as $N$ tends to infinity. 
This has, however, been proven for sufficiently small values of $\b$: 
more precisely, one knows that

\theo{1.1.}{\it Define $\tilde\b_2=1$, and for $p>2$
$$
\tilde\b^{2}_p\equiv \inf_{0<m<1}(1+m^{-p})\phi(m)
\Eq(1.6)
$$
where
$
\phi(m)\equiv [(1-m)\ln (1-m) +(1+m)\ln (1+m)]/2.
$
Then for all $\b< \tilde\b_p$
$$
\lim_{N\uparrow +\infty} F_{\b,N,p} =-\b/2.
\Eq(1.8)
$$
}
\remark For $p=2$ this result was first proven in [ALR]. A very simple proof
has later been given by Talagrand [T]. Comets [C] has shown that the value
$\b=1$ is optimal in the sense that \eqv(1.8) fails for $\b>1$. The result
for $p\geq 3$ is due to Talagrand [T1]. It is clear that in all cases
\eqv(1.8) will fail for $\b\geq\sqrt{2\ln 2}$ which by a more elaborate 
computation can be improved to $\b \geq \sqrt{2\ln 2}(1-2^{-c_pp})$
with $c_p<5$, for $p$ large [B2]. On the other hand, 
a simple calculation 
shows that $\tilde\b_p\sim \sqrt{2\ln 2}(1-2^{-p}/2\ln 2)$. 
One should note that
to get \eqv(1.8) up to a value so close to $\sqrt{2\ln 2}$
required a substantial modification of the original argument of [T2], 
namely the use of a ``truncated'' second moment method. Such a truncation will
also be the main difficulty in obtaining our results\note{For similar reasons,
slightly different truncations were also used by Toubol  [Tou] (and probably 
first) in the
study of the CLT for the SK model with vector valued spins.}.

In the case of the REM, it is well known that the critical inverse
temperature $\tilde{\b}_{REM}=\sqrt {2\ln 2}$ and that
[D2]
$$
\lim_{N\uparrow +\infty} F_{\b,N,REM} =\cases  -\b/2,& \,\hbox{if}
\,\,\b\leq \sqrt{2\ln2}\cr
-\sqrt{2\ln 2} +\b^{-1}\ln 2,
& \,\hbox{if}
\,\,\b\geq \sqrt{2\ln2}.
\endcases
\Eq(1.9)
$$
As a consequence one has that (this result is essentially contained
in [D1], a rigorous proof follows easily from the results 
contained in [T1]\note{Private communication by M. Talagrand.})
$$
\lim_{p\uparrow +\infty} \limsup_{N\uparrow +
\infty} F_{\b,N,p}=\lim_{p\uparrow +\infty} \liminf_{N\uparrow +
\infty} F_{\b,N,p}=\lim_{N\uparrow +\infty} F_{\b,N,REM}.
\Eq(1.10)
$$
In this note we will control the fluctuations of the free energy in 
(essentially) all of the domain of parameters $\b$, $p$ (even)
 where the limit is 
known to exists, i.e. the high temperature regions of the 
$p$-spin 
models, and the {\it entire} temperature range in the REM. Although the
REM is rather singular and the techniques used for that case are totally 
different from those we will use for the $p$-spin models, we felt it
would be instructive to include this singular limiting case in this  paper.
Moreover, it turns out that in spite of the heavy investigation the REM has 
enjoyed over the years [D1,D2,OP,GMP,Ru], no precise fluctuation results
for the free energy are available in the literature. Finally, we are 
convinced that the reader will be rather surprised by the rich structure
the fluctuation behaviour this model exhibits.

Let us now state our results. We begin with the $p$-spin SK models. 

\theo{1.2} {\it Consider the $p$-spin SK-model with 
 $p=2k\geq2$.
   There exists $\b_p>0$ such that for all  $\b<\b_p$
$$
  N^{(p-2)/4} \ln \frac{Z_{\b,N}}{\E Z_{\b, N}}\limlaw
     M_{\infty}(\sqrt \b)
\Eq(1.12)
$$
in distribution as $N\uparrow +\infty$,   
where $M_{\infty}(t)$ is the centered Gaussian process
 with mean zero and covariance
  $$ 
\E(M_{\infty}(t)-M_{\infty}(s))^{2}=(t-s)\big[(p-1)!!\big].
\Eq(1.13)
$$
}
The value of $\b_p$ can be estimated reasonably well. To state lower bound on
$\b_p$ we need, however, some notation. We define the functions
$$
\eqalign{
I(m_1,m_2, m_3)
=\frac{1}{4}&\Bigl((1+m_1+m_2+m_3)
\ln(1+m_1+m_2+m_3)\cr
&\ {}+(1-m_1-m_2+m_3)\ln (1-m_1-m_2+m_3)\cr
&\ {}+(1+m_1-m_2-m_3)\ln (1+m_1-m_2-m_3)\cr
&\ {}+(1-m_1+m_2-m_3)\ln (1-m_1+m_2-m_3)\Bigr),\cr
}
\Eq(1.14)
$$
$$
\eqalign{
S_p(m_1,m_2,m_3)
=\Bigl[&
 \Bigl(1+\frac{m_1^p-m_2^p m_3^p}{1-m_3^{2p}}\Bigr)^2
+\Bigl(1+\frac{m_2^p-m_1^p m_3^p}{1-m_3^{2p}}
\Bigr)^2\cr
&\ \ {}+2m_3^p \Bigl(1+\frac{m_1^{p}-m_2^{p} m_3^{p}}{1-m_3^{2p}}\Bigr)
\Bigl(1+\frac{m_2^{p}-m_1^p m_3^{p}}{1-m_3^{2p}}\Bigr)\Bigr]^{1/2},\cr  
}
\Eq(1.15)
$$
$$
R_p(m_1,m_2,m_3)=\frac{2m_1^p m_2^p
m_3^p-m_1^{2p}-m_2^{2p}}{2(1-m_3^{2p})},
\Eq(1.16)
$$
and
$$
\eqalign{
U_p(m_1, m_2, m_3)=&I(m_1, m_2, m_3)(1+m_3^p)
 \Bigl[S_p(m_1, m_2,m_3)\sqrt{2+2m_3^p}\cr
&\quad\qquad\qquad{}+R_p(m_1, m_2, m_3)(1+m_3^p)-(2+m_3^p)\Bigr]^{-1}
}
\Eq(1.17)
$$
on the set
$$
\eqalign{
{\cal A}\equiv&\bigl\{m_1,m_2,m_3\in [-1,1]^{3}\mid
  1-m_1-m_2+m_3>0,
  1-m_1+m_2-m_3>0,\cr
 &\quad\qquad\qquad\qquad\qquad\qquad\qquad\qquad\qquad\qquad\qquad  1+m_1-m_2-m_3>0\bigr\}.
}
\Eq(1.18)
$$
 Note that the function $I(m_1, m_2, m_3)$ is symmetric in $m_1$, $m_2$ and
$m_3$, 
 and that $S_p(m_1, m_2, m_3)$, $R_p(m_1, m_2, m_3)$ and $U_p(m_1, m_2,
m_3)$ 
 are symmetric in $m_1$ and $m_2$.
  Let
$$
\eqalign{
Y_p(m_1,m_2, m_3)
=&
\max\Bigl\{  
    I(m_1, m_2, m_3)\left(
\frac{2}{3}+\frac{1}{m^{p}_1+m^{p}_2+m^{p}_3}\right),\cr 
&\qquad\quad\;\ U_p(m_1, m_2, m_3),\ U_p(m_1, m_3, m_2),\ U_p(m_2, m_3,
m_1)\Bigr\}.
}
\Eq(1.19)
$$
With this notation we have 

\theo{1.3}{\it 
Let $p=2k>2$.
Then 
$$
\inf_{m_1, m_2, m_3 \in \cal A} Y_p(m_1, m_2, m_3)\leq \b^2_p< {2\ln 2}.
\Eq(1.20)
$$
In particular
$$
\lim_{p\uparrow +\infty} \b_p^2=2\ln 2.
\Eq(1.21)
$$
}
We see that the scale on which the partition functions fluctuate decreases
rapidly as $p$ increases. One might guess that the scale becomes 
exponentially small in $N$ in the limiting random energy model.
This is indeed true, but more surprising things happen, as the following 
theorem states:

\theo {1.4}{\it The free energy of the REM has the following 
fluctuations:
\item{(i)} If $\b<\sqrt{\ln 2/2}$, then 
$$
e^{\frac N2(\ln 2-\b^2)}\ln \frac {Z_{\b,N}}{\E Z_{\b,N}} \limlaw
\NN(0,1).
\Eq(1.22)
$$
\item{(ii)} If $\beta=\sqrt{\ln 2/2}$,  then 
 $$ \sqrt{2} 
 e^{\frac N2(\ln 2-\b^2)}\ln \frac {Z_{\b,N}}{\E Z_{\b,N}} \limlaw
\NN(0,1).
\Eq(1.22bis)
$$       
\item{(iii)} Let $\alpha\equiv\beta/{\sqrt{2\ln 2}}$. 
             If $ \sqrt{\ln 2/2}<\b<\sqrt{2\ln 2}$, then
$$
e^{\frac N2(\sqrt{2\ln 2}-\b)^2+\frac\a 2[\ln (N\ln 2)+\ln 4\pi]} \ln 
\frac {Z_{\b,N}}{\E Z_{\b,N}} \limlaw \int_{-\infty}^\infty
e^{\a z}(\PP(dz) - e^{-z}dz),   
\Eq(1.23)
$$
where $\PP$ denotes the Poisson point process on $\RR$ with intensity 
measure $e^{-x}dx$.
}

Theorem 1.3 covers the high temperature regime. However, in the REM we can 
also compute the fluctuations in the low temperature phase. 

\theo{1.5} {\it  
\item{(i)}If $\beta=\sqrt{2\ln 2}$, then 
$$ 
e^{\frac 12[\ln(N\ln 2)+\ln 4\pi]}\Bigl(
\frac{Z_{\b,N}}{ \E Z_{\b,N}}-\frac{1}{2}+\frac {\ln(N\ln 2)+\ln 4\pi}{4\sqrt{\pi N\ln 2}}
\Bigr)
\!\limlaw\!  \int_{-\infty}^0
e^{\a z}(\PP(dz) - e^{-z}dz)   + \int\limits_{0}^\infty e^{ z} \PP(dz). 
\Eq(1.24)
$$  
\item{(ii)} If
$\b>\sqrt{2\ln 2}$, then 
$$
e^{-N[\b \sqrt{2\ln 2}-\ln 2]+\frac{\a}2[\ln (N\ln 2)+\ln 4\pi]}
Z_{\b,N}  \limlaw \int\limits_{-\infty}^\infty e^{\a z} \PP(dz)
\Eq(1.25)
$$
and
$$
\ln Z_{\b,N}-\E \ln Z_{\b,N}\limlaw  
\ln \int\limits_{-\infty}^\infty e^{\a z} \PP(dz)
-\E \ln \int\limits_{-\infty}^\infty e^{\a z} \PP(dz).
\Eq(1.25a)
$$
}
\remark Note that  expressions like  $ \int_{-\infty}^0
e^{\a z}(\PP(dz) - e^{-z}dz)$ are always understood as
$
\lim_{y\downarrow -\infty}  \int_{y}^0
e^{\a z}(\PP(dz) - e^{-z}dz).
$
We will see that all the functionals of the Poisson point process
appearing are almost surely finite random variables.

\remark Note that the Poisson integral  
$\int_{-\infty}^\infty e^{\a z} \PP(dz)$ is the partition function of 
Ruelle's version of the REM [Ru]. Thus \eqv(1.25a) affirms that above the 
critical temperature, the fluctuations of the free energy of the 
REM converge in distribution to those of Ruelle's model. While this connection 
was surely evident for Ruelle and motivated the introduction of his model,
we have not been able to find a rigorous statement of this connection
in the literature. In [GMP] the scale on which fluctuations take place
has been established, but no actual limit theorem was proven.

\remark It is interesting to observe that in the REM 
 there is a second ``phase 
transition'' within the high-temperature phase at which the fluctuations
become non-Gaussian. In fact, in the REM the main phase transition can
be interpreted as a breakdown of the Law of Large Numbers, while the
second transition corresponds to a breakdown of the Central Limit Theorem. 

The remainder of this paper is organized as follows. In the next section 
we present the proofs of Theorems 1.2 and 1.3. They are based on an adaptation 
of the martingale method of Comets and Neveu. The essential new ingredient 
is  the rather involved truncation procedure inspired by Talagrand's work.
However, in the proof of the CLT, the computational aspects become even 
more involved and require the consideration of truncated third moment
of the partition function. For this reason Section~2 is rather long and
quite technical. However, the proof is organized in such a way that the 
CLT is first proven for ``very high'' temperatures where no truncations
are necessary, while the more technical aspects needed to approach the 
critical temperature are dealt with separately later. Section~3
is devoted to proving Theorems 1.4 and 1.5 for the REM. It is technically
completely different and independent from Section~2. It can therefore be
read independently from the rest of the paper.
In an appendix we explain some of the technical difficulties that
appear in the case $p$ odd and we explain the result to be expected in that
case.     

\bigskip      
     
\chap{2. The CLT in the $p$-spin model}2

The proof of the central limit theorem in the $p$-spin SK model 
relies on a martingale central limit theorem which uses that fact that
a Gaussian random variable can always be seen as the marginal distribution
of a Brownian motion. Thus we follow Comets and Neveu and introduce 
the $p$-parameter family of independent standard Brownian motions 
$(J_{i_1, i_2,\ldots, i_p}(t), t\in {\R}^{+})_{i_1,i_2,\ldots, i_p\in {
\N}}$ with 
$\E J_{i_1,i_2,\ldots,i_p}(t)=0$ and $\E J^{2}_{i_1, i_2,\ldots, i_p}(t)= t$.
 The Hamiltonian of the $p$-spin SK model
 can then be written as $H_N(\s, t)=\sqrt N X_\s(t)$, where 
$$
X_\s(t)=\frac{1}{\sqrt{N^{p}}}\sum_{1\leq i_1,i_2,\ldots,i_p\leq N}
 J_{i_1,i_2,\ldots,i_p}(t)\s_{i_1}\s_{i_2}\cdots\s_{i_p}.
\Eq(2.1)  
$$
Note that we can also consider it as a Gaussian process on $\{-1,1\}^N\times
\R^+$ with mean zero and correlation function 
$$
\hbox{\rm cov}\, (X_\s(t), X_{\s'}(s))= (s\land t) \,
f_p\Big(R_N(\s,\s')\Big),
\Eq(2.2)
  $$
    where $f_p(x)=x^{p}$.
In particular, we have $\E H_N^{2}(\s,t)= Nt$ and
    $\E\exp\{H_N(t,\s)\}=\exp\{Nt/2\}$  for all $\s$.
For later convenience we introduce the {\it normalized} partition function  
$$
  \bar{Z}_{N}(t)=
\E_{\s}\exp\{H_N(t,\s)-Nt/2\},
\Eq(2.3)
  $$
It is  related to the partition function $Z_{\beta,N}$ of Section 1 by
$\bar{Z}_N(\beta^{2})=Z_{\b,N}/\E Z_{\b,N}$, with equality holding in law.
   The important point of this construction is that
  that  for all fixed $N>1$,
   $\bar{Z}_N(t)$ is a {\it continuous martingale} in the variable
   $t$  with $\E \bar{Z}_N(t)=1$.

We begin the proof with some preliminary steps  
    along the lines of~\cite{CN}.
  Let us find the bracket $<\bar{Z}_N(t)>$  of the martingale $\bar{Z}_N(t)$,
  i.\ e.\ the unique increasing process vanishing at zero,
   such that
   $\bar{Z}^{2}_{N}(t)-<\bar{Z}_N(t)>$ is the continuous martingale (see~\cite{RY}).
   By Ito's formula,  $\bar{Z}_N(t)$
 satisfies the following stochastic differential
 equation:
$$
    \hbox{\rm d} \bar{Z}_N(t)=\E_{\s} \exp\{H_N(t,\s)-Nt/2\}\,\hbox{\rm d} H_N(t,\s).
\Eq(2.5)
  $$
 Then due to  well-known  properties
  of  martingale brackets
 $$\eqalign{
<\bar{Z}_N(t)>
=& \E_{\s, \s'} < \int\limits_{0}^{t}
  \!\!e^{H_N(s,\s)-Ns/2}\,\hbox{\rm d} H_{N}(s,\s),
\int\limits_{0}^{t}\!\! e^{H_N(s,\s')-Ns/2}\,\hbox{\rm d}H_{N}(s,\s')>\cr
 =& \E_{\s,\s'} \int\limits_{0}^{t}
 e^{H_N(s,\s)+
 H_N(s,\s')-Ns}\, \hbox{\rm d}<H_{N}(s,\s), H_N(s, \s')>\cr
 =& \E_{\s,\s'} \int\limits_{0}^{t}
e^{H_N(s,\s)
 +H_N(s,\s')-Ns}\,
 N f_p\Big(R_N(\s,\s')\Big)\, \hbox{\rm d} s.
}
\Eq(brz)  $$
    Since
$$
\eqalign{
\E \int\limits_{0}^{t} \bar{Z}_N^{-2}(s)\,\hbox{\rm d}<\bar{Z}_N(s)>
     =& \E\int\limits _{0}^{t} \frac
{\E_{\s, \s'}
  e^{H_N(s,\s)+
H_N(s,\s')-Ns}\,
 N f_p\Big(R_N(\s,\s')\Big)}
{\E_{\s, \s'}
  e^{H_N(s,\s)
+H_N(s,\s')-Ns}}\,
 \hbox{\rm d} s
\leq N t<\infty,
}
\Eq(sq)  $$
  we may introduce a continuous local martingale
$
M_N(t)=\int_{0}^{t} \bar{Z}_N^{-1}(s)\, \hbox{\rm d} \bar{Z}_N(s).
$
  Thus  $\bar{Z}_N(t)$ solves the stochastic differential equation
 $$
\hbox{\rm d}\bar{Z}_N(t)= \bar{Z}_N (t)\,\hbox{\rm d}M_N(t)
$$    
and  the following fundamental
 representation of $\bar{Z}_N(t)$ holds:
$$
 \bar{Z}_N(t)=\exp\{M_N(t)-1/2<M_N(t)>\}.
\Eq(reprz) 
$$
  Here $<M_N(t)>$ is the bracket of $M_N(t)$ and
     $
  <M_N(t)>=\int_0^{t} \bar{Z}_N^{-2}(s)\,\hbox{\rm d}<\bar{Z}_N(s)>.  
$
   Let us note that
$$
\eqalign{
\frac{\hbox{\rm d}}{\hbox{\rm d}t} <M_N(t)>\,=\,&
\bar{Z}_N^{-2}(t)\frac{ d}{dt} <\bar{Z}_N(t)>\cr
 \,=\,& \bar{Z}_N^{-2}(t) \bigg(\E_{\s,\s'}
e^{H_N(t,\s)+H_N(t,\s')
-Nt}\,
 N f_p\Big(R_N(\s,\s')\Big)\bigg).
}
\Eq(2.100)  
$$
      Note also that $M_N(t)$ is locally square integrable.
  In fact, by~\eqv(sq)
$$
\E M_N^{2}(t)= \E<M_N(t)>=\E\int\limits_0^{t} \bar{Z}_N^{-2}(s) \,\hbox{\rm d}
<\bar{Z}_N(s)>\leq N t<\infty.
\Eq(2.1000)  $$

 To prove Theorems 1.2 and 1.3, we will show that
 for all $t$ satisfying 
$$
t<\inf_{m_1, m_2, m_3 \in \cal A} Y_p(m_1, m_2, m_3).
\Eq(tinf)
$$
the bracket
  of the local martingale~$N^{(p-2)/4}M_N(t)$, which is
 $N^{(p-2)/2}<M_N(t)>$, converges
 to~$t \E\xi^{p}$ in probability as $N\uparrow + \infty$.
Here $\xi$ is a Gaussian random variable with $\E\xi=0$, $\E \xi^2=1.$
 Then by the martingale convergence theorem
 (see Theorem~3.1.8 in~\cite{JS}) the local martingale~$N^{(p-2)/4} M_N(t)$
 converges to~$M_{\infty}(t)$ in law as $N\uparrow +\infty$.
 This fact together with the representation~\eqv(reprz)
  implies immediately the statement of Theorem~1.2.

\medskip

\noindent{\bf Sketch of the proof of Theorems 1.2 and 1.3:} 
 We will now  outline further steps of the proof.
 First, we show the convergence  $N^{(p-2)/2}<M_N(t)>\rightarrow t\E \xi^p$
 on a more restricted interval of $t$.
 Lemma~\thv(lemma1) reduces this problem  to the convergence of
  $$
  N^{(p-2)/2}\E |V_N(t)| \rightarrow 0, \qquad \,\hbox{as }N\uparrow +
\infty,
\Eq(cc1)
 $$
  where 
$$V_N(t):=
N^{-\frac{p-2}2}\E_{\s, \s'} \Big(N^{p/2} f_p\Big(R_N(\s,\s')\Big)-
\E\xi^p\Big)
e^{H_N(t,\s)+H_N(t, \s')-Nt}.$$
  The proof  of this lemma is based 
on the fact that 
$$
N^{(p-2)/2}\frac{\hbox{\rm d}}{\hbox{\rm d}t} <M_N(t)>- \E
\xi^{p}=N^{(p-2)/2}\frac{V_N(t)}{\bar{Z}_N^2(t)},  
\Eq(cc2)
  $$
and is performed via integration. It almost mimics the proof proposed
in~\cite{CN}.
In particular, we use the fact that $\bar{Z}_N^{2}(t)$ is not small
on  events of large probability.
    The convergence~\eqv(cc1) is proved in Proposition~\thv(vwt).
Let us give some intuition for it.  One can write 
$$
 \E V_N(t)=\!\!\!\!\sum\limits_{m=0, \pm 1/N, \ldots, \pm 1}\!\!\!\!   
    (N f_p(m)-N^{(2-p)/2} \E \xi^p)e^{tN f_p(m)}
\P(\s\cdot\s'=m N).
\Eq(c3) 
$$
  By Stirling's formula 
$$ \P(\s\cdot\s'=m N) \sim \frac{2}{\sqrt{2\pi (1+m)(1-m) N}}
 e^{-N \phi(m)},
$$
 where  $\phi (m)=[(1+m)\ln (1+m)+(1-m)\ln (1-m)]/2.$
(here and in the sequel we use the symbol $\sim$ to denote asymptotic
equivalence, i.e. $a_N\sim b_N \Leftrightarrow \lim_{N\uparrow +\infty} 
\frac {a_N}{b_N}=1$).
   Note that $\phi(m)=-m^{2}/2(1+o(1))$ as $m\rightarrow 0$.
 Now split the right-hand side of~\eqv(c3) into two terms:
 the summation  in the first one will be  over $m$ with $|m|$ ''small
enough''
   and in the second --- over all other $m$.
  It is not difficult to treat the first term. Since $p\geq 3$,
we have for any fixed $t$ 
$$
tf_p(m)+\phi(m)=-m^{2}/2(1+o(1)),\qquad m\rightarrow 0.
\Eq(c') 
 $$
   Then  putting $m\sqrt{N}=s$, the first term  becomes
$$
\frac{2}{\sqrt{2\pi N}}\sum\limits_{s= m\sqrt {N}}   
    (N^{(2-p)/2}s^{p} - N^{(2-p)/2} \E \xi^p)e^{-s^{2}/2}
\sim \frac{2 N^{(2-p)/2}}{\sqrt{2\pi}}
\int\limits_{-\infty}^{\infty}(s^{p}-\E \xi^{p})
e^{-s^2/2}\, \hbox{\rm d}s,
$$
   from where the normalisation $N^{(p-2)/2}$ is immediate.
To ensure the convergence to zero of the second term (the one with
correlations
  $m$ not close to zero), 
  the power of the exponent in it should be
 negative: 
$$ \sup_{m\in [0,1]} (tf_{p}(m)- \phi(m))<0.$$ 
   Thus for all  $t<\inf_{0<m<1} \phi(m)m^{-p}$, 
 we get
$ N^{(p-2)/2} \E V_N(t) \rightarrow 0.$
  Note that,  Proposition~\thv(vwt) states a 
  stronger result~\eqv(cc1).
  To get rid of the absolute value of $V_N(t)$ in~\eqv(cc1), 
  we follow an idea suggested in~\cite{CN}
to  apply the Cauchy-Schwartz inequality.
  Thus, instead of $\E |V_N(t)|$, we get $W_N(t)$  
  (see the proof of Proposition~\thv(vwt))
  which refers to the third moment of $\bar{Z}_N(t)$. 
   This makes technical computations slightly 
   tougher and leads to the bound on $t$  \eqv(tinf2) given in Lemma 
     \thv(lemma1) below.
        
   Note also that these arguments are valid only for $p\geq 3$.
   The case $p=2$  of~\cite{CN} and [Tou] is different, since there, 
   \eqv(c') does not hold. This case is treated in~\cite{CN}
   by the multi-dimensional  Central Limit Theorem  for 
  $N$ independent vectors $(\s_i\s'_i, \s'_i\s''_i, \s_i\s''_i)$.

     Next, we  will extend the bound~\eqv(tinf2) to the full regime
    announced in~\eqv(tinf).
 We have seen, that~\eqv(tinf2)  was imposed 
 by configurations of spins with rather big correlations~$m$
  in the sum~\eqv(c3).
  We will reduce their contribution,  using  Talagrand's idea 
   to truncate the Hamiltonian.        
 Consider instead of $V_N(t)$ 
$$
\eqalign{
\widetilde{V}_N(t,\epsilon)
=&\E_{\s, \s'}\Big(N f_p \Big(R_N(\s,\s')\Big)- 
N^{(2-p)/2}\E\xi^p\Big)
e^{H_N(t,\s)+H_N(t,\s')-Nt}\cr
&\qquad\qquad\qquad\qquad\qquad\qquad\qquad\qquad{}
\times \1_{\{H_N(t,\s)<(1+\epsilon)tN,
  H_N(t,\s')<(1+\epsilon)tN\}}
}
$$
  for some $\epsilon>0$.  Then
$$
\eqalign{
\E \widetilde{V}_N(t, \epsilon)=&\sum\limits_{m=0, \pm 1/N, \ldots, \pm 1}
  (N f_{p}(m)- tN^{(2-p)/2}) \P(\s\cdot\s'= mN)\cr
&\qquad\qquad\qquad{}\times \E e^{\sqrt{Nt}\xi_1+\sqrt{Nt}\xi_2-Nt}
\1_{\{\xi_1<\sqrt{Nt}(1+\epsilon), \xi_2<\sqrt{Nt}(1+\epsilon)\}},
}
\Eq(ccc1)
 $$  
where $\xi_1, \xi_2$ are standard Gaussians with 
$\hbox{\rm cov}\, (\xi_1, \xi_2)=m$.
   Let us again split $\E \widetilde{V}_N(t, \epsilon)$  into two terms 
 with ''small'' and ''large'' $m$ in the sum~\eqv(ccc1).
 The analysis of the first term is completely analogous 
 to the one in the case of~$V_N(t)$.    
 We can neglect the truncation here,
  since $\xi_1$ and $\xi_2$ are almost  independent.
    In the second term, $\xi_1$ and  $\xi_2$ are  more correlated. 
 But due to the truncation, the expectation of the exponent involved in this
term is much smaller than $e^{tm^p N}$. 
 In fact, by the
elementary estimate \eqv(stand1)  for Gaussian random variables
$$\eqalign{
\E & e^{\sqrt{Nt}\xi_1+\sqrt{Nt}\xi_2-Nt} 
\1_{\{\xi_1<\sqrt{Nt}(1+\epsilon), \xi_2<\sqrt{Nt}(1+\epsilon)\}}\cr 
& \leq  
\E e^{\sqrt{Nt(2+2m ^{p})}\xi-Nt}
\1_{\{ \xi<2 \sqrt{Nt}(1+\epsilon)(2+2m^p)^{-1}\}}\cr
\vphantom{\int} {} & \leq 
\exp\{[-4 Nt(1+\epsilon)^2][4+4m^p]^{-1}+2Nt(1+\epsilon)-Nt\} \cr
& = \exp\{[Ntm^{p}(1+2\epsilon)-Nt \epsilon^2][1+m^p]^{-1}\}.
}$$      
   Then for any 
$$
 t <\inf_{0<m<1} (1+m^{-p})\phi(m)
\Eq(c6) 
 $$
  and for an appropriate choice of $\epsilon$
 all terms of the sum~\eqv(ccc1) with  $m$ not  close 
  to zero  are exponentially small.
  This implies $N^{(p-2)/2}\E \widetilde{V}_N(t, \epsilon) \rightarrow 0$. 
The bound~\eqv(c6) is  Talagrand's bound for the critical temperature 
in the $p$-spin SK model, see~\eqv(1.6).  It tends to $2\ln2$ as
$p\uparrow + \infty$.

  In order to incorporate this idea into our proof,
   we reduce the problem of convergence 
  $N^{(p-2)/2}<M_N(t)>\rightarrow t \E \xi^p$ 
   to  the  following statements:
$$
N^{(p-2)/2} \E |\widetilde{V}_N(t, \epsilon)| \rightarrow  0,
\Eq(c7)
$$
and
$$
 N^{(p-2)/2} \E | (V_N(t)-\widetilde{V}_N(t, \epsilon))\bar{Z}^{-2}_N(t)| 
\rightarrow  0,
\Eq(c8)
  $$
  for all $\epsilon>0$.
  This is derived in Lemma~\thv(le2) again 
   from~\eqv(cc1).   
In Proposition~\thv(prp2) we show \eqv(c7).
 Again, because of the absolute value,  
   we must  apply 
the Cauchy-Schwartz inequality and  pass to the third 
 moment of~$\bar{Z}_N(t)$. 
   This makes technical computations  much harder.
 Namely, we get three standard Gaussian 
 random variables $\xi_1, \xi_2, \xi_3$ with  
 covariances $m_1$, $m_2$, $m_3$.
 To benefit from the truncation for obtaining a good bound on~$t$,
 we have to take into account four different cases:
 one when all  $m_1, m_2, m_3$ are large and 
 others when two of these correlations are large
 and the third  is small. Then the analogue of \eqv(c6) 
 is the minimum of four estimates of this kind. 
 Therefore,  the bound~\eqv(tinf) is  the minimum 
 of  four  functions. 
     The convergence \eqv(c8)  is the subject of Proposition~\thv(prp3).
  Its proof uses ideas of M.~Talagrand and 
 a concentration of measure inequality.

\lemma{\TH(lemma1)}{\it
  Let
$$
 T<\inf_{\cal A}\frac{I(m_1, m_2, m_3)}{m_1^p+m_2^p+m_3^p}.
\Eq(tinf2)
  $$
   Then
$$
\sup\limits_{0\leq t\leq T}
|N^{(p-2)/2}<M_N(t)>-t \E\xi^{p}| \rightarrow 0
\Eq(lem1) 
 $$
in probability, where
$\xi$ is a Gaussian random variable with
$\E\xi=0$, $\E\xi^2=1$.
}

\noindent{\bf Proof.}
   Let us denote by
$$V_N(t)=\frac{\hbox{\rm d}}{\hbox{\rm d}t} <\bar{Z}_N(t)>-N^{(2-p)/2} \bar{Z}_N^{2}(t)\E\xi^{p}.$$
   Then
$$\eqalign{
\frac{\hbox{\rm d}}{\hbox{\rm d}t} N^{(p-2)/2}<M_N(t)>- \E\xi^{p} 
=& N^{(p-2)/2}V_N(t)\bar{Z}_N^{-2}(t)\cr
=& N^{(p-2)/2}V_N(t)\exp\{-2M_N(t)+<M_N(t)>\}.
}
$$    Let us introduce the events
$$ A^{N}_{a,b}:=\{-M_N(t)\leq a +(b/2)<M_N(t)>\ \,\hbox{ for all }t\geq 0\}.$$
   Note that by an appropriate choice of~$a>0$ and~$b>0$,
their probabilities can be made arbitrarily close to~$1$.
In fact, the process $B_N(t)=M_N(S_{t})$, where
$S_t=\min\{s\mid <M_N(s)>=t\}$, is a standard Brownian motion
  and $M_N(t)=B_N(<M_N(t)>)$.
  By the well-known fact for Brownian motion
$$
 \P\{A^N_{a,b}\}=\P\{-B_N(t)\leq a+(b/2)t\ \,\hbox{ for all }t\geq 0\}
\geq 1-\exp\{-ab\}.
\Eq(events) 
 $$
    We  have:
$$
\eqalign{
\Big|
\Big(N^{(p-2)/2} &\frac{\hbox{\rm d}}{\hbox{\rm d}t}<M_N(t)>-\E\xi^{p}\Big)
\1_{\{A^N_{a,b}\}}\Big|\cr
\vphantom{\int\limits_{a}^{b}}\,=\,& N^{(p-2)/2}|V_N(t)|\exp\{-2M_N(t)+<M_N(t)>\} \1_{\{A^N_{a,b}\}}\cr
\,\leq\, & N^{(p-2)/2}\exp\{2a\}|V_N(t)|\exp\{(1+b)<M_N(t)>\}.
}\Eq(vr2)
 $$
   Let us also introduce the function
 $\chi_b(x):=[1-\exp\{(1+b)x\}][1+b]^{-1}.$
   Then by~\eqv(vr2) for all $t\leq T$
$$
\eqalign{
|N^{(p-2)/2}& \chi_b(<M_N(t)>-t N^{(2-p)/2}
\E\xi^{p}) \1_{\{A^N_{a,b}\}}|\cr
=&N^{(p-2)/2}\Big|\int\limits_{0}^{t}
   \Big(\frac{\hbox{\rm d}}{\hbox{\rm d}s} <M_N(s)>- N^{(2-p)/2}\E\xi^{p}\Big)
\1_{\{A^N_{a,b}\}}\cr
  &\qquad\qquad\quad\quad{}\times\exp\{-(1+b)(<M_N(s)>-sN^{(2-p)/2}\E\xi^{p})\}
\hbox{\rm d} s\Big|
}
$$
$$
\eqalign{
\quad\leq&
N^{(p-2)/2}\int\limits_{0}^{t}
 \Big|\frac{\hbox{\rm d}}{\hbox{\rm d}s} <M_N(s)>- N^{(2-p)/2} \E\xi^{p}\Big|
  \1_{\{A^N_{a,b}\}}\cr
&\qquad\qquad\qquad\qquad{}\times\exp\{-(1+b)(<M_N(s)>-s N^{(2-p)/2}\E\xi^p)\}\,
\hbox{\rm d} s \cr
\quad\leq & N^{(p-2)/2} \exp\{2a+T N^{(2-p)/2}(1+b)\}\int\limits_{0}^{t}
|V_{N}(s)| \,\hbox{\rm d}s.
}
\Eq(2.101)  
$$
  This yields
$$
\eqalign{
N^{(p-2)/2}&\sup\limits_{0\leq t\leq T}
|\chi_{b}(<M_{N}(t)>- t N^{(2-p)/2} \E\xi^p)| \1_{\{A^N_{a,b}\}}
\cr
\leq& N^{(p-2)/2}
 \exp\{2a+T N^{(2-p)/2}(1+b)\}
\int\limits_{0}^{T}|V_N(s)|\,\hbox{\rm d}s.
}\Eq(2.102)  
$$
  We will show in Proposition~\thv(vwt)  that 
  $$\lim\limits_{N\uparrow +\infty} N^{(p-2)/2}\E|V_N(t)|=0$$
  uniformly in $t\in[0, T]$. Consequently
   $ \sup_{N>1,t\leq T} N^{(p-2)/2}\E|V_N(t)|<\infty.$
 Then by the dominated convergence theorem
$$\lim_{N\uparrow +\infty}
   \E\big[N^{(p-2)/2} \sup_{0\leq t\leq T}
    |\chi_{b}(<M_N(t)>-t N^{(2-p)/2}\E\xi^p)|
     \1_{\{A^N_{a,b}\}}\big]=0.$$
      It follows that for all $a,b>0$
   $$N^{(p-2)/2} \sup_{0\leq t\leq T}
    |\chi_{b}(<M_N(t)>-t N^{(2-p)/2}\E\xi^p)|
     \1_{\{A^N_{a,b}\}}\rightarrow 0 \quad \,\hbox{ as } N\uparrow + \infty$$ 
   in probability. Then also
    $N^{(p-2)/2} \sup_{0\leq t\leq T}
    |\chi_{b}(<M_N(t)>-t N^{(2-p)/2}\E\xi^p)|\to \infty,$ as $N \to \infty$
    since by~\eqv(events) the probability of the events
    $A^N_{a,b}$ can be made arbitrarily close to~$1$. 
     This last fact implies~\eqv(lem1) and the lemma is proved.\endproof

     It remains to prove the following proposition.

\proposition{
\TH(vwt)}{\it
 Assume that $T$ satisfies~\eqv(tinf2).
  Then
$$
\lim\limits_{N\uparrow +\infty}N^{(p-2)/2}\E| V_N(t)|=0
\Eq(pr1)  $$
 uniformly in $[0, T]$.
}

\noindent {\bf Proof.}
  It follows from~\eqv(brz) and the definition of $V_N(t)$ that
$$\eqalign{
N^{(p-2)/2}V_N(t)
=&
\E_{\s, \s'} \Big(N^{p/2} f_p\Big(R_N(\s,\s')\Big)-\E\xi^p\Big)
e^{H_N(t,\s)+H_N(t, \s')-Nt}.
}
$$
  By the Cauchy-Schwartz inequality:
$$\eqalign{
N^{(p-2)/2}\E|V_N(t)|
=&\E\Big|\E_{\s}e^{H_N(t,\s)-Nt/2}\E_{\s'}
\Big(N^{p/2}f_p\Big(R_N(\s,\s')\Big)-\E\xi^p\Big)
e^{H_N(t,\s')-Nt/2}\Big|\cr
\leq &
\Big[ \E \E_{\s} e^{H_N(t,\s)- Nt/2}\Big]^{1/2}\cr
 &{}\times \Big[\E \E_{\s} e^{H_N(t,\s)-Nt/2}
\Big[\E_{\s'}
\Big(N^{p/2} f_p\Big(R_N(\s,\s')\Big)-\E\xi^p\Big)
e^{H_N(t,\s')-Nt/2}\Big]^{2}\Big]^{1/2}\cr
=&{}\Big[\E_{\s,\s',\s''}
\Big(N^{p/2}f_p\Big(R_N(\s,\s')\Big)-\E\xi^p\Big)
\Big(N^{p/2}f_p\Big(R_N(\s,\s'')\Big)-\E\xi^p\Big)\cr
&{}\times \exp\Big\{Nt\Big(f_p\Big(R_N(\s,\s')\Big)
+f_p\Big(R_N(\s,\s'')\Big)+f_p
\Big(R_N(\s(\s',\s'')\Big)\Big)
\Big\}\Big]^{1/2}.
}
$$
Then it suffices to prove that
$$\eqalign{
W_N(t)=&\E_{\s,\s',\s''}
\Big(N^{p/2}f_p\Big(R_N(\s,\s')\Big)-\E\xi^p\Big)
\Big(N^{p/2}f_p\Big(R_N(\s,\s'')\Big)-\E\xi^p\Big)\cr
&\qquad\quad{}\times \exp\Big\{Nt\Big(f_p\Big(R_N(\s,\s')\Big)
+f_p\Big(R_N(\s,\s'')\Big)+f_p
\Big(R_N(\s(\s',\s'')\Big)\Big)\Big\}
}
$$
 tends to zero uniformly
  in $[0,T]$ as $N\uparrow +\infty$. 
 We  represent it as
$$\eqalign{
W_N(t)=&\!\!\!\!\!\!\!\!\!\sum\limits_{m_1,m_2,m_3 \in {\cal A}_N}
\!\!\!\!\!\!\!\!\!\Big(N^{p/2}f_p(m_1)-\E\xi^p\Big)
\Big(N^{p/2}f_p(m_2)-\E\xi^p\Big)
e^{Nt(f_p(m_1)
+f_p(m_2)+f_p(m_3))}\cr
&\qquad\qquad\quad\times \P\{\s\cdot\s'=m_1N,\ \s\cdot\s''= m_2 N,\
   \s'\cdot\s''=m_3 N\}
}
$$
where the set ${\cal A}_N={\cal A}\cap\{0,\pm 1/N,\pm 2/N,\ldots,\pm
1\}^{3}$.
A standard combinatorial calculation yields
$$
\eqalign{
&\P\{\s\cdot\s'=m_1N, \s\cdot\s''=m_2N,
\s'\cdot\s''=m_3N\}\cr
 & = 2^{-2N}
   {N\choose N(1+m_1)/2}
  {N(1+m_1)/2\choose N(1+m_1+m_2+m_3)/4}
{N(1-m_1)/2\choose N(1+m_2-m_1-m_3)/4}.
}
\Eq(I)  
$$
By Stirling's formula we obtain
$$
\eqalign{
\P&\{\s\cdot\s'=m_1N, \s\cdot\s''=m_2N,
\s'\cdot\s''=m_3N\}\cr
&=\frac{16 \exp\{-N I(m_1,m_2,m_3)\}}
   {\sqrt{(2\pi)^3 N^3}}
 [(1+m_1+m_2+m_3)(1-m_1-m_2+m_3)]^{-1/2}
\cr
&\qquad{}\times [(1+m_1-m_2-m_2)(1-m_1+m_2-m_3)]^{-1/2}
\Big(1+O\Big(\frac{1}{N}\Big)\Big)\ \,\hbox{as
}N\uparrow +\infty,
}
\Eq(prst)
 $$
 for any given $m_1,m_2,m_3 \in{\cal A}_N$.  Let us remark that
$$
t(m_1^{p}+m_2^p+m_3^p)+(m_1^2+m_2^2+m_3^2)/2-I(m_1,m_2,m_3)
=O\big((|m_1|+|m_2|+|m_3|)^{3}\big)
\Eq(Iexp)
$$
 as $m_1, m_2, m_3\rightarrow 0$ uniformly in $[0, T]$. 
Then for all sufficiently small $\epsilon>0$
 there exists a constant $h>0$ such that
$$
\sup\limits_{t\in[0,T]}
\big[ t(m_1^p+m_2^p+m_3^p)-
 I(m_1,m_2,m_3) \big]<-h(m_1^2+m_2^2+m_3^2)/2
\Eq(h)
  $$
 for all $m_1, m_2, m_3\in {\cal A}\cap\{|m_1|+|m_2|+|m_3|<\epsilon\}$.
   Let us fix such a small $\epsilon>0$
  and an arbitrary constant $0<\delta<1/6$
 and then split $W_N(t)$ into four terms:
$$W_N(t)=I^{1}_{N}+I^{2}_{N}(t)+I^{3}_{N}(t)+I^{4}_{N}(t),$$
where
$$
\eqalign{
I^{1}_{N}=&\frac{16}{\sqrt{(2\pi N)^3}}\!\!\!\!\!\!\!\!\!\!\!
\sum\limits_{m_1,m_2,m_3\in {\cal A}_N \atop
 |m_1|+|m_2|+|m_3|<N^{-1/3-\delta}}
\!\!\!\!\!\!\!\!\!\!\!\!\!\!\!\!\Big(N^{p/2}f_p(m_1)-\E\xi^p\Big)
\Big(N^{p/2}f_p(m_2)-\E\xi^p\Big)e^{-N(m_1^2+m_2^2+m_3^2)/2}\cr
\vphantom{\sum\limits^{A}}
I^{2}_{N}(t)=&\!\!\!\!\!\!\!\!\!\!
\sum\limits_{m_1,m_2,m_3\in {\cal A}_N \atop\!\!\!\!\!
 |m_1|+|m_2|+|m_3|<N^{-1/3-\delta}}^{}
 \!\!\!\!\!\!\!\!\!\!\!\!\!\Big(N^{p/2}f_p(m_1)-\E\xi^p\Big)
\Big(N^{p/2}f_p(m_2)-\E\xi^p\Big)\cr
&\qquad {}\times\Big(e^{Nt(f_p(m_1)
+f_p(m_2)+f_p(m_3))}\P(\s\cdot\s'=m_1N,\ \s\cdot\s''= m_2 N,\
  \s'\cdot\s''=m_3 N)\cr
&\qquad\qquad\qquad\qquad\qquad\qquad\qquad\qquad\qquad\qquad\ \ \ \ \ \ {}-\frac{16}{\sqrt{(2\pi
N)^3}}e^{-N(m_1^2+m_2^2+m_3^2)/2}\Big),\cr
I^{3}_{N}(t)
=&\!\!\!\!\!\!\!\!\!\!
\sum\limits_{{m_1,m_2,m_3\in {\cal A}_N
\atop |m_1|+|m_2|+|m_3|>N^{-1/3-\delta}}
 \atop |m_1|+|m_2|+|m_2|<{\scriptstyle \epsilon}}
 \!\!\!\!\!\!\!\!\!\!\!\!\!\!\Big(N^{p/2}f_p(m_1)-\E\xi^p\Big)
\Big(N^{p/2}f_p(m_2)-\E\xi^p\Big)e^{Nt(f_p(m_1)
+f_p(m_2)+f_p(m_3))}\cr
&\qquad\qquad\qquad\qquad\qquad\quad\quad\ {}\times \P(\s\cdot\s'=m_1N,\ \s\cdot\s''= m_2 N,\
   \s'\cdot\s''=m_3 N),\cr
\vphantom{\sum\limits^{A}}I^{4}_{N}(t)
=&
\!\!\!\!\!\!\!\!\!\!\!\sum\limits_{m_1,m_2,m_3\in {\cal A}_N
\atop |m_1|+|m_2|+|m_3|>{\scriptstyle \epsilon}}
 \Big(N^{p/2}f_p(m_1)-\E\xi^p\Big)
\Big(N^{p/2}f_p(m_2)-\E\xi^p\Big)e^{Nt(f_p(m_1)
+f_p(m_2)+f_p(m_3))}\cr
&\;\qquad\qquad\qquad\qquad\quad\quad\qquad\quad\ \times \P(\s\cdot\s'=m_1N,\ \s\cdot\s''= m_2
N,\
   \s'\cdot\s''=m_3 N).
}
$$
 We will prove that all four terms $I^{1}_{N}, I^{2}_{N}(t), 
 I^{3}_{N}(t), I^{4}_{N}(t)$
 tend to zero uniformly in $[0,T]$
 as $N\uparrow + \infty$.

 To show this  for $I^{1}_{N}$, let us put
$m_1\sqrt N=s_1, m_2\sqrt N=s_2, m_3\sqrt N=s_3$. Then
$$
\eqalign{
\lim_{N\uparrow +\infty}I^{1}_{N}
=&\lim_{N\uparrow +\infty}\frac{16}{\sqrt{(2\pi N)^3}}\!\!\!\!\!\!\!\
\sum\limits_{{s_1,s_2,s_3\atop
\qquad\ \ \ \ =0,\pm 1/\sqrt{N},
\pm 2/\sqrt{N},\ldots}\atop
\ \ \ |s_1|+|s_2|+|s_3|<N^{1/6-\delta}}
\!\!\!\!\!\!\!\!\!\!\!\!\!\!\!\!(s_1^p-\E\xi^p)(s^2_p-\E\xi^p)e^{-(s_1^2+s_2^2
+s_3^2)/2}\cr
=& \frac{16}{\sqrt{(2\pi)^{3}}}
\int\limits_{-\infty}^{\infty}
\int\limits_{-\infty}^{\infty}
\int\limits_{-\infty}^{\infty}
(x^p- \E\xi^p)(y^p-\E\xi^p) e^{-(x^2+y^2+z^2)/2}\hbox{\rm d} x \hbox{\rm d} y
 \hbox{\rm d}
z
=0.
}
$$

 To treat $I^{2}_{N}(t)$, we rewrite it using~\eqv(prst) as
$$
\eqalign{
I^{2}_{N}(t)
=&
\frac{16}{\sqrt{(2\pi N)^3}}\!\!\!\!\!\!\!\!\!\!\!\!\!\!
\sum\limits_{m_1,m_2,m_3\in {\cal A}_N \atop
 |m_1|+|m_2|+|m_3|<N^{-1/3-\delta}}
\!\!\!\!\!\!\!\!\!\!\!\!\!\!\! \Big(N^{p/2}f_p(m_1)-\E\xi^p\Big)
\Big(N^{p/2}f_p(m_2)-\E\xi^p\Big)
e^{-N(m_1^2+m_2^2+m_3^2)/2}\cr
&\qquad\qquad\ {}\times\Big[e^{N [t(f_p(m_1)+
+f_p(m_2)+f_p(m_3))+(m_1^2+m_2^2+m_3^2)/2-I(m_1,m_2,m_3)]}\cr
&\qquad\qquad\quad\quad {}\times
 [(1+m_1+m_2+m_3)(1-m_1-m_2+m_3)]^{-1/2}\cr
&\qquad\qquad\quad\quad
{}\times[(1+m_1-m_2-m_3)(1-m_1+m_2-m_3)]^{-1/2}
\Big(1+O\Big(\frac{1}{N}\Big)\Big)-
1\Big].
}\Eq(2.106)  $$
   Moreover, here  O(1) is 
bounded uniformly  in ${\cal A}_N\cap\{|m_1|+|m_2|+|m_3|<\epsilon\}$
  by Stirling's formula. 
     It follows from~\eqv(Iexp) that
$$
\lim_{N\uparrow +\infty}\!\!\!\!\!\!\!\!\!\!
\sup\limits_{{t\in[0, T]}\atop
|m_1|+|m_2|+|m_3|<N^{-1/3-\delta}}\!\!\!\!\!\!\!\!\!\!\!
N|t(m_1^{p}+m_2^p+m_3^p)+(m_1^2+m_2^2+m_3^2)/2-I(m_1,m_2,m_3)|=0.
$$
  Then
$$\eqalign{
&\lim_{N\uparrow +\infty}\!\!\!\!\!\!\!\!\sup\limits_
{{t\in[0, T] }\atop
|m_1|+|m_2|+|m_3|<N^{-1/3-\delta}}
\!\!\!\!\!\!\!\!\!\!\!\!\!\!\!\!\!\!\!\!
\Big|e^{N[t(f_p(m_1)
+f_p(m_2)+f_p(m_3))
+(m_1^2+m_2^2+m_3^2)/2-I(m_1,m_2,m_3)]}\cr
&\quad\qquad\quad\qquad\quad\quad {}\times
 [(1+m_1+m_2+m_3)(1-m_1-m_2+m_3)]^{-1/2}\cr
&\quad\qquad\quad\qquad\quad\quad{}
\times[(1+m_1-m_2-m_2)(1-m_1+m_2-m_3)]^{-1/2}
\Big(1+O\Big(\frac{1}{N}\Big)\Big)-1
\Big|=0,
}
$$
  while
$$
\eqalign{
\lim_{N\uparrow +\infty}&\frac{16}{\sqrt{(2\pi N)^3}}\!\!\!\!\!\!\!\!\!\!
\sum\limits_{m_1,m_2,m_3\in {\cal A}_N \atop
 |m_1|+|m_2|+|m_3|<N^{-1/3-\delta}}
\!\!\!\!\!\!\!\!\!\!\!\! \Big|\Big(N^{p/2}f_p(m_1)-\E\xi^p\Big)
\Big(N^{p/2}f_p(m_2)-\E\xi^p\Big)\Big|e^{-N(m_1^2
+m_2^2+m_
3^2)/2}\cr
=&\lim_{N\uparrow +\infty}
\frac{16}{\sqrt{(2\pi N)^3}}
\sum\limits_{
  {s_1,s_2,s_3 =0,\pm 1/\sqrt{N},\ldots} \atop
 |s_1|+|s_2|+|s_3|<N^{1/6-\delta}}\!\!\!\!\!
 \Big|\Big(s_1^p-\E\xi^p\Big)
\Big(s_2^p-\E\xi^p\Big)\Big| e^{-(s_1^2+s_2^2+s_3^2)/2}\cr
=& \frac{16}{\sqrt{(2\pi)^{3}}}
\int\limits_{-\infty}^{\infty}
\int\limits_{-\infty}^{\infty}
\int\limits_{-\infty}^{\infty}
|(x^p- \E\xi^p)(y^p-\E\xi^p)| e^{-(x^2+y^2+z^2)/2}\hbox{\rm d} x \hbox{\rm d} y \hbox{\rm
d} z<\infty.
}
$$
  Thus $I^{2}_{N}(t)\rightarrow 0$ uniformly in $[0,T]$  as 
$N\uparrow +\infty$.

   To estimate $I^{3}_{N}(t)$, we rewrite it in the same way
using~\eqv(prst): 
$$
\eqalign{
I^{3}_{N}(t)=&
\frac{16}{\sqrt{(2\pi N)^3}}\!\!\!\!\!\!\!\!
\sum\limits_{{m_1,m_2,m_3\in {\cal A}_N \atop
 |m_1|+|m_2|+|m_3|>N^{-1/3-\delta}}\atop
  |m_1|+|m_2|+|m_3|<\epsilon}
\!\!\!\!\!\!\!\! \Big(N^{p/2}f_p(m_1)-\E\xi^p\Big)
\Big(N^{p/2}f_p(m_2)-\E\xi^p\Big)\cr
&\qquad\qquad\qquad\qquad\qquad
e^{N[t(f_p(m_1)+f_p(m_2)+f_p(m_3))-I(m_1,m_2,m_3)]}\cr
&\qquad\qquad\qquad\ {}\times
 [(1+m_1+m_2+m_3)(1-m_1-m_2+m_3)]^{-1/2}\cr
&\qquad\qquad \qquad {}\times[(1+m_1-m_2-m_2)(1-m_1+m_2-m_3)]^{-1/2}
\Big(1+O\Big(\frac{1}{N}\Big)\Big)
}
\Eq(2.107)  
$$
  Due to~\eqv(h),
there exists a constant $h'>0$ such that for all sufficiently large~$N$
$$
\eqalign{
&\sup\limits_{
{t\in[0,T]
\atop |m_1|+|m_2|+|m_3|>N^{-1/3-\delta}}
\atop {|m_1|+|m_2|+|m_3|<\epsilon}}
\exp\{-N[t(m_1^p+m_2^p+m_3^p)-I(m_1,m_2,m_3)]\} \cr
&\leq
\sup_
{|m_1|+|m_2|+|m_3|>N^{-1/3-\delta}\atop
 |m_1|+|m_2|+|m_3|<\epsilon}
 \exp\{-N h (m_1^2+m_2^2+m_3^2)/2\}
\leq \exp\{-h'N^{1/3-2\delta}\}.
}
$$
The sum
$$
\sum\limits_{
{m_1,m_2,m_3\in {\cal A}_N \atop
 |m_1|+|m_2|+|m_3|>N^{-1/3-\delta}}
\atop
  |m_1|+|m_2|+|m_3|<\epsilon}
\!\!\!\!\!\! \Big(N^{p/2}f_p(m_1)-\E\xi^p\Big)
\Big(N^{p/2}f_p(m_2)-\E\xi^p\Big)
$$
has  polynomial growth as $N\uparrow +\infty$ and
 the uniform convergence $I^{3}_{N}(t)\rightarrow 0$
   in $[0, T]$ is proved.
 
Finally, let us consider $I^{4}_{N}(t)$. By Stirling's
formula
there exists a constant $C$ such that
 for all $(m_1, m_2, m_3)\in {\cal A}_N\cap\{|m_1|+|m_2|+|m_3|>\epsilon\}$
$$
\P\{\s\cdot\s'=m_1N, \s\cdot\s''=m_2N, \s' \cdot\s''=m_3 N\}
 \!\leq \!C\sqrt{N}\!\exp\{-N I(m_1,m_2,m_3)\}.
\Eq(bp) 
 $$
  Then by the assumption~\eqv(tinf2), for given~$T$
 there exists a constant  $h''>0$ such that
$$
\eqalign{
&\sup_{
t\in[0,T] 
\atop |m_1|+|m_2|+|m_3|>\epsilon
}
\exp\{Nt(m_1^p+m_2^p+m_3^p)\}  
\P\{\s\cdot\s'=m_1N, \s\cdot\s''=m_2N, \s'\cdot s''=m_3 N\}\cr
\leq& C\sqrt{N}\!\!\!\!\!\!\!\!\!\!\!\!\!\!
\sup_{t\in[0,T] 
\atop |m_1|+|m_2|+|m_3|>\epsilon}
\exp\{-N[t(m_1^p+m_2^p+m_3^p)-I(m_1,m_2,m_3)]\}
<C\sqrt{N}\exp
\{-h''N\}.
}
$$
 The remaining sum in this term
  has again  polynomial growth, whence
  $I^{4}_{N}(t)\rightarrow 0$ uniformly in $[0, T]$.
  The lemma is proved.\endproof

\medskip

\noindent{\bf Remark.}  Let us note that the restriction~\eqv(tinf2) 
 on~$T$ was essential only  for the analysis of the  fourth term $I^4_N(t)$.
 This means that the convergence $N^{(p-2)/2}\E|V_{N}(t)|\rightarrow 0$ 
  breaks down for larger~$T$ only because of the configurations of spins
  with rather big correlations $\s\cdot\s'=m_1$, 
   $\s\cdot\s''=m_2$, $\s'\cdot\s''=m_3$.    
   To extend our result to the whole interval~\eqv(tinf) of admissible~$T$, 
   we need to reduce the contribution of these configurations
  into~$W_N(t)$.
   For that purpose we will follow the idea of M.~Talagrand~\cite{T}
   to truncate the Hamiltonian.

    Now we prove the statement of the previous
 lemma for all~$T$ satisfying~\eqv(tinf).

\lemma{\TH(le2)}{\it
Let
$$
 T<\inf_{m_1, m_2,m_3\in {\cal A}}Y(m_1,m_2,m_3).
$$
 Then
$$
\sup\limits_{0\leq t\leq T}
 |<N^{(p-2)/4} M_N(t)>-t\E\xi^p|\rightarrow 0
\Eq(conv2)  $$
in probability.
}

\noindent{\bf Proof.}  
   Let us fix  $\epsilon>0$ such that for some constants 
    $h_1, h_2>0$
$$
\sup_{t\in[0, T] \atop m_1^p+m_2^p+m_3^p<3\epsilon} \!\!\!\!\!\!\!
\big[ t(m_1^p+m_2^p+m_3^p)-I(m_1,m_2,m_3)\big]
<-h_1(m_1^2+m_2^2+m_3^2)
\Eq(c1)
  $$
and 
$$
\eqalign{
&\sup_{{t\in[0,T] \atop m_1, m_2, m_3 \in {\cal A}}\atop
 m_1^p+m_2^p+m_3^p>3\epsilon}\!\!\! t\big[\min\,
\{Q_p(m_1, m_2, m_3, \epsilon),\ L_p(m_1, m_2, m_3, \epsilon), \cr
&\qquad\quad\quad\qquad\quad L_p(m_1, m_3, m_2, \epsilon),\  L_p(m_2, m_3, m_1,
\epsilon)\}\big]-I(m_1, m_2, m_3)<-h_2 
}
\Eq(c2)  
$$ 
 where
$$\eqalign{
Q_p(m_1, m_2, m_3, \epsilon)
 =&[-9\epsilon^2+6(1+2\epsilon)(m_1^p+m_2^p+m_3^p)]
[ 2(3+2m_1^p+2m_2^p+2m_3^p)]^{-1},\cr
L_p(m_1, m_2, m_3, \epsilon)
=&\Big[-1-m_3^p-(1+\epsilon)^2+(1+\epsilon)S_p(m_1, m_2, m_3)
\sqrt{2+2 m_3^p}\cr
&\qquad\qquad\qquad\qquad\qquad\qquad\ {}+R_p(m_1, m_2,
m_3)(1+m_3^p)\Big][1+m_3^p]^{-1}.
}
$$  
    Condition~\eqv(c1) is the same as \eqv(h) and,
   due to~\eqv(Iexp), for any given $T>0$ it is possible to find
 an appropriate $\epsilon>0$ such that~\eqv(c2) is satisfied.  
However,
 $\epsilon>0$  ensuring~\eqv(c2) does exist, 
  if and only if~ $T$ satisfies
   the assumption~\eqv(tinf).
  The meaning of~\eqv(c2) will become clear in the proof
   of a further Proposition~\thv(prp2). 
   Let us introduce
$$
\eqalign{
\widetilde{V}_N(t,\epsilon)
=&\E_{\s, \s'}\Big(N f_p \Big(
R_N(\s,\s')\Big)- N^{(2-p)/2}\E\xi^p\Big)
e^{H_N(t,\s)+H_N(t,\s')-Nt}\cr 
&\qquad\qquad{}\times \1_{\{H_N(t,\s)<(1+\epsilon)tN,
  H_N(t,\s')<(1+\epsilon)tN\}}\cr
\bar{V}_N(t,\epsilon)=&
\E_{\s, \s'}\Big(N f_p \Big(
R_N(\s,\s')\Big)- N^{(2-p)/2}\E\xi^p\Big)
e^{H_N(t,\s)+H_N(t,\s')-Nt}\cr
&\qquad\qquad{}\times \1_{\{H_N(t,\s)>(1+\epsilon)tN,
\,\hbox{ or }H_N(t,\s')>(1+\epsilon)tN\}}\cr
=&V_N(t)-\widetilde{V}_N(t, \epsilon).
}
$$
  Let  us also fix some
$T_0>0$ satisfying the assumption~\eqv(tinf2)
   of the previous lemma.
 Proceeding along the  lines
  of the proof of Lemma~\thv(lemma1),
  we get for all $t\in[T_0, T]$:
$$
\eqalign{
&N^{(p-2)/2}
|F_{b}(<M_{N}(t)>- tN^{(2-p)/2} \E\xi^p)| \1_{\{A^N_{a,b}\}}\cr
&\leq N^{(p-2)/2}
 \exp\{2a+T_0 N^{(2-p)/2}(1+b)\}
\int\limits_{0}^{T_0}|V_N(s)|\,\hbox{\rm d}s\cr
&\quad{}+ N^{(p-2)/2}
 \exp\{2a+ t N^{(2-p)/2}(1+b)\}
\int\limits_{T_0}^{t}|\widetilde V_N(s, \epsilon)|\,\hbox{\rm d}s\cr
&\quad{}+N^{(p-2)/2}\int\limits_{T_0}^{t}|\bar{V}_N(s,\epsilon)|\bar{Z}_N^{-2}(s)
\exp\{-(1+b)(<M_N(s)>-s N^{(2-p)/2})\}
\1_{\{A^N_{a,\epsilon}\}}\, \hbox{\rm d}s.
}
$$
   Then 
$$
\eqalign{
&N^{(p-2)/2} \sup\limits_{T_0<t\leq T}
|F_{b}(<M_{N}(t)>- tN^{(2-p)/2} \E\xi^p)| \1_{\{A^N_{a,b}\}}\cr
&\leq N^{(p-2)/2}
 \exp\{2a+T_0 N^{(2-p)/2}(1+b)\}
\int\limits_{0}^{T_0}|V_N(s)|\,\hbox{\rm d}s\cr
&\quad{}+ N^{(p-2)/2}
 \exp\{2a+ T N^{(2-p)/2}(1+b)\}
\int\limits_{T_0}^{T}|\widetilde V_N(s,\epsilon)|\,\hbox{\rm d}s\cr
&\quad{}+N^{(p-2)/2}\exp\{ TN^{(2-p)/2}(1+b)\}
\int\limits_{T_0}^{T}|\bar{V}_N(s,\epsilon)|\bar{Z}_N^{-2}(s)\hbox{\rm d}s. 
}
$$
   It was proved in Lemma~\thv(lemma1) that
$N^{(p-2)/2}\E|V_N(t)|\rightarrow 0$  
 uniformly in $[0, T_0]$ as $N\uparrow + \infty$. 
 Proposition~\thv(prp2)  shows that for $\epsilon>0$ 
  satisfying~\eqv(c1) and~\eqv(c2),
    $N^{(p-2)/2}\E|\widetilde{V}(t, \epsilon)|\rightarrow 0$ 
  uniformly in $t\in[T_0, T]$.
    Proposition~\thv(prp3) proves that 
    $N^{(p-2)/2}\E |\bar{V}(t, \epsilon)Z^{-2}_N(t)| \rightarrow 0$
   uniformly in $[T_0, T]$ for all $\epsilon>0$.
Then 
$$\lim_{N\uparrow + \infty}
\E[\sup_{T_0\leq t\leq T}
|N^{(p-2)/2}  F_b(<M_N(t)>-t N^{(2-p)/2}
\E\xi^{p})| \1_{\{A^N_{a,b}\}}]=0.$$ 
  Then~$\sup_{0\leq t\leq T}|N^{(p-2)/2} F_b(<M_N(t)>-t N^{(2-p)/2}
\E\xi^{p})|$ converges to zero in probability, 
since the probability of the events $A^N_{a,b}$ can be made 
  arbitrarily close to~$1$ by~\eqv(events). This
  implies~\eqv(conv2) and the proof of the lemma is complete.

\proposition{\TH(prp2)}{\it
  Assume that~$T>0$ satisfies~\eqv(tinf).
   Let us fix~$0<\epsilon<1/2$  such that
  \eqv(c1) and \eqv(c2) hold. Then 
  for any $T_0>0$, $T_0<T$:
$$
\lim_{N\uparrow + \infty} N^{(p-2)/2}\E |\widetilde{V}_N(t,\epsilon)|=0
\Eq(eq1)
  $$
uniformly in $t\in[T_0, T]$.
}

\noindent {\bf Proof.}          
  Let us estimate $N^{(p-2)/2}\E |\widetilde{V}_N(t,\epsilon)|$
 by the Cauchy-Schwartz inequality as   in the  proof 
 of Proposition~\thv(vwt) for $N^{(p-2)/2}\E|V_N(t)|$.
 After that we split it into four terms:
$$
\eqalign{
 N^{(p-2)/2} \E|\widetilde{V}_N(t, \epsilon)|\leq &
   [\E\widetilde{W}_N(t, \epsilon)]^{1/2}
=[\widetilde{I}^{1}_{N}(t,\epsilon)-\widetilde{I}^{2}_{N}
(t,\epsilon)
 +\widetilde{I}^{3}_{N}(t,\epsilon)+\widetilde{I}^{4}_{N}
(t,\epsilon)]^{1/2},
}
$$
where 
$$
\eqalign{
\widetilde{W}_N(t,\epsilon)
=&\E \E_{\s, \s', \s''}
\Big(N^{p/2}f_p\Big(R_N(\s,\s')\Big)-\E\xi^p\Big)
\Big(N^{p/2}f_p\Big(R_N(\s,\s'')\Big)-\E\xi^p\Big)\cr
&\quad\qquad\quad{}\times e^{H_N(t, \s)+H_N(t, \s')+H_N(t, 
\s'')-3tN/2}\cr
&\quad\qquad\quad{}\times \1_{\{
  H_N(t,\s)<Nt(1+\epsilon), H_N(t,\s')<Nt(1+\epsilon),H_N(t,
\s'')<Nt(1+\epsilon)\}}
}
$$
$$
\eqalign{
\widetilde{I}^{1}_{N}(t,\epsilon)=&\!\!
\!\!\!\!\!\!\sum\limits_{m_1,m_2,m_3 \in {\cal A}_N
\atop m_1^p+m_2^p+m_3^p\leq \epsilon^2/4}
 \!\!\!\!\!\!\Big(N f_p(m_1)-N^{(2-p)/2}\E\xi^p\Big)
\Big(N f_p(m_2)-N^{(2-p)/2}\E\xi^p\Big)\cr
&\qquad\quad\times \P\{\s\cdot\s'=m_1N,\ \s\cdot\s''= m_2 N,\
   \s'\cdot\s''=m_3 N\}\cr
&\qquad\quad{}\times\E e^{H_N(t,\s)+H_N(t,\s')+H_N(t,\s'')-3tN/2}
}
$$
$$
\eqalign{
\widetilde{I}^2_N(t, \epsilon)=&
\!\!\!\!\!\!\!\!\!\sum\limits_{m_1,m_2,m_3 \in {\cal A}_N
 \atop m_1^p+m_2^p+m_3^p <\epsilon^2/4}
 \!\!\!\!\!\!\!\Big(Nf_p(m_1)- N^{(p-2)/2}\E\xi^p\Big)
\Big(N f_p(m_2)- N^{(p-2)/2} \E\xi^p\Big)\cr
&\qquad\quad\times \P\{\s\cdot\s'=m_1N,\ \s\cdot\s''= m_2 N,\
   \s'\cdot\s''=m_3 N\}\cr
&\qquad\quad{}\times \E [e^{H_N(t, \s)+H_N(t, \s')+H_N(t, \s'')-3Nt/2}\cr
&\qquad\quad\ \ {}\times
\1_{\{H_N(t,\s)\geq N t(1+\epsilon)\;\,\hbox{or}\;H_N(t,\s') \geq
Nt(1+\epsilon),\,\hbox{ or }
H_N(t,\s'') \geq Nt (1+\epsilon)\}}]\cr
\widetilde{I}^{3}_{N}(t,
\epsilon)=&\!\!\!\!\!\!\!\!\!\sum\limits_{m_1,m_2,m_3 \in {\cal A}_N
 \atop \epsilon^2/4\leq m_1^p+m_2^p+m_3^p \leq  3\epsilon}
\!\!\!\!\!\!\!\!\!\!\!\Big(N f_p(m_1)- N^{(2-p)/2}\E\xi^p\Big)
\Big(N f_p(m_2)-N^{(2-p)/2}\E\xi^p\Big)\cr
&\qquad\quad{}\times \P\{\s\cdot\s'=m_1N,\ \s\cdot\s''= m_2 N,\
   \s'\cdot\s''=m_3 N\}\cr
&\qquad\quad{}\times \E [e^{H_N(t, \s)
+H_N(t, \s')+H_N(t, \s'')-3Nt/2}\cr
&\ \qquad\quad\ \ {}\times
\1_{\{H_N(t,\s)< N t(1+\epsilon), H_N(t,\s')<Nt(1+\epsilon),H_N(t,\s'')<Nt
(1+\epsilon)\}}]\cr
\widetilde{I}^{4}_{N}(t,
\epsilon)=&\!\!\!\!\!\!\!\!\sum\limits_{m_1,m_2,m_3 \in {\cal A}_N
 \atop m_1^p+m_2^p+m_3^p >3\epsilon}
 \!\!\!\!\!\!\!\Big(Nf_p(m_1)- N^{(p-2)/2}\E\xi^p\Big)
\Big(N f_p(m_2)- N^{(p-2)/2} \E\xi^p\Big)\cr
&\qquad\quad\times \P\{\s\cdot\s'=m_1N,\ \s\cdot\s''= m_2 N,\
   \s'\cdot\s''=m_3 N\}\cr
&\qquad\quad{}\times \E [e^{H_N(t,\s)+
H_N(t, \s')+H_N(t, \s'')-3tN/2}\cr
&\qquad\quad\ \ {}\times
\1_{\{H_N(t,\s)\geq N t(1+\epsilon)\;\,\hbox{or}\;H_N(t,\s') \geq
Nt(1+\epsilon)\,\hbox{ or }   H_N(t,\s'')
\geq Nt (1+\epsilon)\}}].\cr
}
$$
  We will prove the uniform convergence  to zero in $[T_0, T]$ 
  as $N\uparrow + \infty$  of all these four terms.

  The first term $\widetilde{I}^1_N(t)$ is not truncated and it refers 
  to the configurations of spins with small correlations
  $m_1$, $m_2$ and $m_3$.
 The proof of its uniform convergence to zero in~$[T_0, T]$  
 relies on~\eqv(c1)
 and it is completely analogous to the proof of the uniform convergence 
 to zero of the sum $I^1_N+I^2_N(t)+I^3_N(t)$ in the proof of Proposition~1.
Therefore, 
 we omit the details.

  The second term $\widetilde{I}^2_N(t)$ also contains
  only configurations of spins  with very small correlations.
  If these correlations were zero, i.\ e.\ if $H_N(t,\s)$ 
   $H_N(t, \s')$ and $H_N(t, \s'')$ were independent, then, indeed,
   the expectation involved in this term    satisfies 
$$\E [e^{H_N(t,\s)+H_N(t, \s')+H_N(t, \s'')-3Nt/2}\1_{\{\cdot\}}]
   \leq 3\E [e^{\sqrt{Nt}\xi-Nt/2} \1_{\{\xi>\sqrt{Nt}(1+\epsilon)\}}]\leq 
\exp\{-Nt\epsilon^2/2\}
$$
 ($\xi$ is a standard Gaussian)  
 by a well-known estimate for Gaussian
  random variables~\eqv(stand).   
   We show that very small correlations~$m_1$,  $m_2$, $m_3$ 
  do not destroy the exponential convergence to zero of 
    the corresponding expectation. 
     Considering the third term~$\widetilde{I}^3_N(t)$, we 
  neglect the truncation and use the asymptotic expansion~\eqv(prst)
  and condition~\eqv(c1). 
  So we prove that the expectation $\E e^{H_N(t, \s)+H_N(t, \s')+H_N(t,
\s'')-3Nt/2}$
   multiplied by the probability of any given correlations
  goes to zero exponentially fast.
  Finally~$\widetilde{I}^4_N(t)$ refers to the configurations 
  of spins with rather big correlations.    
  Here, applying the estimate~\eqv(stand), we benefit from the truncation.
  The choice of $\epsilon>0$ according to~\eqv(c2)
  plays a crucial role in the analysis
  of this term. (Remember that this choice was possible only 
  for~$T$ satisfying~\eqv(tinf)). 
       
     Now we proceed with the detailed proof.    
   To treat the second term $\widetilde{I}^2_N(t, \epsilon)$, we write
$$
\eqalign{
&\E[e^{H_N(t,\s)+H_N(t,\s)+H_N(t,\s'')-3Nt/2}
\!\1_{\{H_N(t,\s)>Nt(1+\epsilon)\hbox{ or}\;H_N(t,\s')>Nt(1+\epsilon)
\;\,\hbox{or}\;H_N(t,\s'')>Nt(1+\epsilon)\}}]\cr
\vphantom{\sum^{A}}&=\E[e^{\sqrt{Nt}(\xi_1+\xi_2+\xi_3)-3Nt/2}
\,\1_{\{\xi_1>\sqrt{Nt}(1+\epsilon)\,\hbox{or}\;
\xi_2>\sqrt{Nt}(1+\epsilon)
\;\,\hbox{or}\;
\xi_3>\sqrt{Nt}(1+\epsilon)\}}],
}
$$
  where $\xi_1$, $\xi_2$ and $\xi_3$ are Gaussian random variables
   with zero mean, variance~$1$ 
  and covariances $\hbox{\rm cov}\, (\xi_1, \xi_2)=
f_p(R_N(\s,\s'))=m_1^p$,
   $\hbox{\rm cov}\, (\xi_1, \xi_3)=f_p(R_N(\s,\s''))=m_2^p$, 
   $\hbox{\rm cov}\, (\xi_2, \xi_3)=f_p(R_N(\s',\s''))=m_3^p$,
   $m_1^p+m_2^p+m_3^p\leq \epsilon^2/4$.
One gets
$$
\eqalign{
\E[ e^{\sqrt{Nt}(\xi_1+\xi_2+\xi_3)-3Nt/2}
\1_{\{\xi_1>\sqrt{Nt}(1+\epsilon)\}}]
\vphantom{\int}=&e^{-3Nt/2} \E \big[
e^{\sqrt{Nt}\xi_1}\1_{\{\xi_1>\sqrt{Nt}(1+\epsilon)\}}\E(
   e^{\xi_2+\xi_3} \mid \xi_1) \big]\cr
=& e^{Nt \gamma-3Nt/2} \E[ e^{
{\sqrt{Nt}(1+\mu)\xi_1}}\1_{\{\xi_1>\sqrt{Nt}(1+\epsilon)\}}],
}
$$  
where $\gamma=1+m_3^p-(m_1^p+m_2^p)^2/2$, $\mu=m_1^p+m_2^p$.
   Since $m_1^p+m_2^p\leq \epsilon^2/4 <\epsilon$, we may use
  the  estimate for standard Gaussian random variables~\eqv(stand). 
  It implies
$$
\eqalign{
\E[ e^{\sqrt{Nt}(\xi_1+\xi_2+\xi_3)-3Nt/2}
\1_{\{\xi_1>\sqrt{Nt}(1+\epsilon)\}}]
\vphantom{\int} \leq& C_1 \exp\{Nt (m_1^p+m_2^p+m_3^p-(\epsilon
-m_1^p+m_2^p)^2/2)\}\cr
\leq & C_1  \exp\{- N T_0\epsilon^2 /8\}
}
$$
  for some constant $C_1>0$, all $t\in[T_0, T]$ and all $N>0$, 
if $m_1^p+m_2^p+m_3^p<\epsilon^2/4$, $0<\epsilon<1/2$.
  Thus
$$
\eqalign{
&\sup_{0\leq m_1^p+m_2^p+m_3^p\leq \epsilon^2/4
}\E [e^{H_N(t,\s)+H_N(t,\s)+H_N(t,\s'')-3Nt/2}\cr
 &\qquad\qquad\qquad\qquad\qquad{}\times \1_{\{H_N(t,\s)>Nt(1+\epsilon)\hbox{ or }H_N(t,\s')>Nt(1+\epsilon)
\hbox{ or }H_N(t,\s'')>Nt(1+\epsilon)\}}]\cr
&\qquad\qquad \leq 3 C_1 \exp\{-N T_0 \epsilon^2/8\}
}
$$  
  for all $t\in[T_0, T]$.
  Since the other terms in~$\widetilde{I}^2_{N}(t, \epsilon)$ 
  have   polynomial growth, the uniform convergence
  $\widetilde{I}^2_{N}(t,\epsilon)\rightarrow 0$ in $[T_0, T]$ 
  follows.

    Let us turn to $\widetilde{I}^3_N(t,\epsilon)$.
    By the expansion~\eqv(prst) and condition~\eqv(c2) 
$$
\eqalign{
&\sup_{\epsilon^2/4\leq m_1^p+m_2^p+m_3^p\leq 3 \epsilon}
\E \bigl[e^{H_N(t, \s)+H_N(t, \s')+H_N(t, \s'')-3Nt/2} \cr
&\qquad\qquad\qquad\qquad\qquad\times
\1_{\{H_N(t,\s)<N t(1+\epsilon),H_N(t,\s')<Nt(1+\epsilon), H_N(t,
\s'')<Nt(1+\epsilon) \}}]\cr
&\qquad\qquad\qquad\qquad \qquad{}\times \P\{\s\cdot\s'=m_1 N,
\s\cdot\s''=m_2 N,
  \s\cdot\s''=m_3 N\}\cr 
&\quad \leq C_2 \sup_{
   \epsilon^2/4\leq m_1^p+m_2^p+m_3^p\leq 3 \epsilon}
\exp\{ N[t(m_1^p+m_2^p+m_3^p)-I(m)] \}\cr
&\quad \leq C_2 
   \sup_{
    m_1^p+m_2^p+m_3^p \geq \epsilon^2/4}
    \exp\{-h_1 N (m_1^2+m_2^2+m_3^2)\}
\leq C_2 \exp\{-h_1 \epsilon^{4/p} N/4\}
}
$$
 for all $t\in[T_0, T]$,  where $C_2>0$, $h_1>0$ are constants.
 All  other terms in $\widetilde{I}^3_{N}(t,\epsilon)$ have  polynomial
   growth, hence $I^3_N(t,\epsilon)\rightarrow 0$    
   uniformly in $[T_0, T]$. 

    Finally,  consider $\widetilde{I}^{4}_N(t, \epsilon)$. 
 We have 
$$
\eqalign{
\E &\left[e^{H_N(t,\s)+H_N(t, \s')+H_N(t, \s'')-3Nt/2 }\,       
 \1_{\{H_N(t, \s)<Nt(1+\epsilon),H_N(t, \s')<Nt(1+\epsilon),
 H_N(t, \s'')<(1+\epsilon)Nt\}}\right]\cr
&\leq\E\left[e^{\sqrt{Nt(3+2m_1^p+2m_2^p+2m_3^p)}\,\xi-3Nt/2} 
\1_{\{\sqrt{3+2m_1^p+2m_2^p+2m_3^p}\,\xi\leq 3Nt(1+\epsilon)\}}\right],
}
$$
 where $\xi$ is a standard Gaussian,
  $m_1=f_p(R_N(\s,\s'))$, $m_2=f_p(R_N(\s,\s''))$, 
  $m_3=f_p(R_N(\s',\s''))$.
  Since $m_1^p+m_2^p+m_3^p>3\epsilon$, we may
 apply the estimate~\eqv(stand1). It yields
$$\eqalign{
\E& \left[e^{H_N(t,\s)+H_N(t, \s')+H_N(t, \s'')-3Nt/2 }\,       
 \1_{\{H_N(t, \s)<Nt(1+\epsilon), H_N(t, \s')<Nt(1+\epsilon),
 H_N(t, \s'')<(1+\epsilon)Nt\}}\right]\cr
&\leq  C_3 \exp\{N t Q_p(m_1, m_2, m_3, \epsilon)\},
}
$$
  for some constant $C_3>0$, all $t\in[T_0, T]$, $N>0$ and
$m_1^p+m_2^p+m_3^p>3\epsilon$.
  On the other hand, we also have:
$$
\eqalign{
\E&\Bigl[e^{H_N(t,\s)+H_N(t, \s')+H_N(t, \s'')-3Nt/2 }\,       
 \1_{\{H_N(t, \s)<Nt(1+\epsilon),H_N(t, \s')<Nt(1+\epsilon),
 H_N(t, \s'')<(1+\epsilon)Nt\}}\Bigr]\cr
&\leq  \E \Bigl[e^{H_N(t,\s)+H_N(t, \s')+H_N(t, \s'')-3Nt/2
}\,       
 \1_{\{H_N(t, \s')<Nt(1+\epsilon),H_N(t, \s'')<Nt(1+\epsilon)\}}\Bigr]\cr
& = \E\Bigl[e^{\sqrt{Nt}\xi_2+\sqrt{Nt}\xi_3-3Nt/2}
\E(e^{\sqrt{Nt}\xi_1} \mid \xi_2, \xi_3)
\1_{\{\xi_2<\sqrt{Nt}(1+\epsilon),
\xi_3<\sqrt{Nt}(1+\epsilon)\}}\Bigr]\cr
&=e^{-Nt+Nt\alpha}\E \Bigl[e^{\sqrt{Nt}(1+\mu_2)\xi_2+
\sqrt{Nt}(1+\mu_3)\xi_3}
\1_{\{\xi_2<\sqrt{Nt}(1+\epsilon),
\xi_3<\sqrt{Nt}(1+\epsilon)\}}\Bigr]\cr
&\leq e^{-Nt+Nt\alpha}
\E \Bigl[e^{\sqrt{Nt((1+\mu_2)^2+(1+\mu_3)^2+2m_3^p(1+\mu_2)(1+\mu_3))}
 \xi}
\1_{\{\sqrt{2+2m_3^p}\, \xi<2\sqrt{Nt}(1+\epsilon)\}}\Bigr], 
}
$$
where $\xi_1$, $\xi_2$, $\xi_3$ are the same as in the analysis
of the second term, $\xi$ is standard Gaussian and
$$
\eqalign{
\alpha=&(2 m_1^pm_2^pm_3^p-m_1^{2p}-m_2^{2p})/(2-2m_3^{2p})\cr
\mu_2=&(m_{1}^{p}-m_2^{p}m_3^{p})/(1-m_3^{2p})\cr
\mu_3=&(m_2^{p}-m_1^{p}m_3^{p})/(1-m_3^{2p}).              
}
$$
   One checks that 
$$
\eqalign{
&\sqrt{(1+\mu_2)^2+(1+\mu_3)^2+2m_3^p(1+\mu_2)
(1+\mu_3)}\cr
&\qquad\geq\frac{
2(1+m_3^p+(m_1^p+m_2^p)/2)}{\sqrt{2+2m_3^p}}
\geq\frac{
  2(1+3\epsilon/2)}{\sqrt{2+2m_3^p}},
}
$$
 when $m_1^p+m_2^p+m_3^p>3\epsilon$.
 So, we are again in the position to apply~\eqv(stand1). This yields
$$\eqalign{
\E&\left[e^{H_N(t,\s)+H_N(t, \s')+H_N(t, \s'')-3Nt/2 }\,       
\1_{\{H_N(t, \s)<Nt(1+\epsilon), H_N(t, \s')<Nt(1+\epsilon),
 H_N(t, \s'')<(1+\epsilon)Nt\}}\right]\cr
&\leq  C_4 \exp\{t N L_p(m_1, m_2, m_3,\epsilon)\},
}
$$   
  where $C_4>0$ is a constant.  
 Permuting  $m_1$, $m_2$ and  $m_3$,
 we  can  derive in the same way that the same expectation 
 does not exceed $\exp\{t N L_p(m_1, m_3, m_2, \epsilon)\}$ and 
 $\exp\{t N L_p(m_2, m_3, m_1, \epsilon)\}$
  multiplied by some constant.
Thus, taking into account~\eqv(bp), we obtain
$$
\eqalign{
&\sup_{m_1^p+m_2^p+m_3^p>3\epsilon}
\E \bigl[e^{H_N(t, \s)+H_N(t, \s')+H_N(t, \s'')-3Nt/2} 
\cr
&\qquad\qquad\qquad\qquad\quad\quad\times
\1_{\{H_N(t,\s)<N t(1+\epsilon),H_N(t,\s')<Nt(1+\epsilon), H_N(t,
\s'')<Nt(1+\epsilon)\}}\bigr] \cr
&\qquad\qquad\qquad\qquad\quad\quad\times\P\{\s\cdot\s'=m_1 N,
\s\cdot\s''=m_2 N,
  \s\cdot\s''=m_3 N\}\cr 
&\leq 
\sup_{m_1^p+m_2^p+m_3^p >3\epsilon}
C_5\sqrt{N} \exp\Bigl\{tN \min \bigl[Q_p(m_1, m_2, m_3, \epsilon), 
L_p(m_1, m_2,m_3, \epsilon),\cr 
&\quad\qquad\qquad\qquad\qquad\qquad\qquad L_p(m_1, m_3, m_2, \epsilon),\;L_p(m_2,m_3,m_1,
\epsilon)\bigr]-NI(m_1, m_2, m_2)\Bigr\}
}
\Eq(las)  
$$
 for all $t\in[0, T_0]$, where $C_5>0$ is a constant.
  Now the relevance of the assumption~\eqv(c2) becomes clear.
Due to~\eqv(c2), the right-hand side of~\eqv(las) tends to zero
exponentially fast, 
 as one can estimate it by $C_5\sqrt{N}\exp\{-h_2N\}$.
 The other terms in~$\widetilde{I}^{4}_{N}(t, \epsilon)$
 have  polynomial growth.
  Thus $I^{4}_N(t, \epsilon)\uparrow +\infty$ uniformly 
  in $[T_0, T]$. This concludes the proof of the proposition.

\proposition{\TH(prp3)}{\it
  For all $T>0$ satisfying \eqv(tinf) and all $\epsilon>0$ 
$$
\lim_{N\uparrow + \infty} N^{(p-2)/2}\E |\bar{V}_N(t,\epsilon)\bar{Z}_{N}^{-2}(t)|=0
\Eq(eq2)
  $$
uniformly in any interval  $[T_0, T]$, where $0<T_0<T$.
}

\noindent{\bf Proof.} It follows from  the definition of $\bar{V}_N(t)$ that
$$
 N^{(p-2)/2}\E |\bar{V}_N(t, \epsilon) \bar{Z}_N^{-2}(t)|\leq \bar{C} N \E
\frac{\E_{\s} e^{H_N(t, \s)}\1_{\{H_N(t,\s)>Nt(1+\epsilon)\}}}
{\E_{\s}e^{H_N(t, \s)}}
\Eq(LK)  
$$
  for all $t\geq 0$, where $\bar{C}>0$ is a constant.
   We will show that the expectation of this last fraction tends 
to zero exponentially fast. First of all, we observe that by~\eqv(stand)
$$
\frac{ \E \E_{\s} e^{H_N(t, \s)} \1_{\{ H_N(t,\s)>Nt(1+\epsilon) \} }}
{ \E \E_{\s} e^{H_N(t, \s)} }
 = 
 \E \E_{\s}e^{H_N(t, \s)-Nt/2} \1_{\{H_N(t,\s)>Nt(1+\epsilon)\}}
\leq
    e^{-Nt\epsilon^2/2}.
\Eq(LK1)
$$
   Let us  represent the fraction in the right-hand side of~\eqv(LK) as
$$
\eqalign{
&
\E\frac{\E_{\s} e^{H_N(t, \s)} \1_{\{H_N(t,\s)>Nt(1+\epsilon)\}} }
{ \E_{\s}e^{H_N(t, \s)} } \cr
&\quad = \E\frac{ 
\E_{\s}e^{H_N(t, \s)-Nt/2}\1_{\{H_N(t,\s)>Nt(1+\epsilon)\}}}
{\exp\{\ln\E_{\s}e^{H_N(t,\s)}-\E\ln\E_{\s} e^{H_N(t, \s)}
+\E\ln\E_{\s}e^{H_N(t, \s)}-Nt/2\}}.
}
\Eq(LK3)  
$$ 
 To expand this formula, we will use the
 concentration of measure as in~\eqv(conc). 
 The random variable $\E_{\s} e^{H_N(t, \s)}$ has the same 
distribution as $\phi(J_1,\ldots, J_{N^p})$, where the function
  $$\phi(x_{1},\ldots, x_{N^p})=\ln \E_\s
\exp\Big\{\sqrt{t N^{1-p}} \sum_{i_1, \ldots, i_p}
x_{i_1,i_2,\ldots, i_p}\s_{i_1}\s_{i_2}\cdots \s_{i_p}\Big\}$$
 is defined on  ${\Z}^{N^p}$, $J_1,\ldots, J_{N^p}$
 are standard Gaussian random variables.
  The Lipschitz constant of $\phi(x_1, \ldots, x_{N^p})$
 is at most  $\sqrt{t N^{1-p}}\sqrt{N^p}=\sqrt{tN}$.
Substituting this function  and $u=Nt\epsilon^2/4$ into~\eqv(conc),
   we derive:
 $$
\P\{|\ln \E_{\s}e^{H_N(t, \s)}-\E\ln \E_{\s}e^{H_N(t, \s)}|
>Nt\epsilon^2/4 \}\leq \exp\{-Nt\epsilon^4/32\}.
\Eq(conc1) 
 $$         
  Let us introduce the events 
$O^N_{t, \epsilon} := \{ |\ln \E_{\s}e^{H_N(t, \s)}-\E\ln
\E_{\s}e^{H_N(t, \s)}|
>Nt\epsilon^2/4\}.$
    Consequently by~\eqv(LK3) and \eqv(conc1)
$$
\eqalign{
 \E &\frac{\E_{\s} e^{H_N(t, \s)} \1_{\{H_N(t,\s)>Nt(1+\epsilon)\}} }
{ \E_{\s}e^{H_N(t, \s)} } \cr
&=\, 
\E\frac{ 
\1_{\{O^{N}_{t,\epsilon}\}}\E_{\s}e^{H_N(t,
\s)-Nt/2}\1_{\{H_N(t,\s)>Nt(1+\epsilon)\}}}
{\exp\{\ln\E_{\s}e^{H_N(t,\s)}-\E\ln\E_{\s} e^{H_N(t, \s)}
+\E\ln\E_{\s}e^{H_N(t, \s)}-Nt/2\}}
+ \P\{O^{N}_{t, \epsilon}\}\cr
& \leq \,
e^{Nt\epsilon^2/4}
\E\frac{ 
\E_{\s}e^{H_N(t, \s)-Nt/2}\1_{\{H_N(t,\s)>Nt(1+\epsilon)\}}}
{\exp\{
\E\ln\E_{\s}e^{H_N(t, \s)}-Nt/2\}}
+ e^{-Nt\epsilon^4/32}
}
\Eq(IL)  $$     
Observe that
  for any $T$ satisfying~\eqv(tinf)  and any $0<T_0<T$, there exists a
constant $K>0$   
  such that
$$
-K\sqrt{N}<\E\ln\E_{\s}e^{H_N(t, \s)}-Nt/2\leq 0
\Eq(est)  
$$   
 for all $t\in[T_0, T]$. The upper bound in~\eqv(est)
 is  immediate by Jensen inequality.
Whenever  the second moment of $\bar{Z}_N(t)$ truncated is finite,
the left-hand side of~\eqv(est) was established by Talagrand~\cite{T1}
in the analysis of the critical temperature.
We will outline his proof in our situation.
 For given $T$ satisfying~\eqv(tinf),
let us fix $\widetilde{\epsilon}>0$, such that~\eqv(c1) 
  and \eqv(c2) hold.  Let us define after
$$\bar{Z}_N(t, \widetilde{\epsilon})= \E_{\s} e^{H_N(t, \s)-Nt/2}
\1_{\{H_N(t, \s)<Nt(1+
\widetilde{\epsilon})\}}
$$
  By~\eqv(stand1) there exists a constant~$K_1<0$ such that 
$$
\E\bar{Z}_N(t, \widetilde{\epsilon})\geq  K_1
\Eq(ex)  
$$
 for all $t\in[T_0, T]$. Moreover, there exists a constant $K_2>0$
   such that 
$$
\E\bar{Z}_N^{3}(t, \widetilde{\epsilon})\leq K_2
\Eq(3exp) 
 $$
for all $t\in[T_0, T]$.
  The proof of~\eqv(3exp) is  analogous to the proof 
  of the uniform convergence to zero of $\widetilde{W}_N(t, \epsilon)$
   in Proposition~2. We decompose
$\bar{Z}_N(t, \widetilde{\epsilon})$ into four terms
  like it was  for $\widetilde W_N(t, \epsilon)$.
 The last three of them go to zero uniformly in~$t\in[T_0, T]$
 and exponentially fast by the same arguments as $\widetilde{I}^{2}_{N}(t)$,
  $\widetilde{I}^{3}_{N}(t)$ and $\widetilde{I}^{4}_{N}(t)$ do.
   We work out the first term similarly to the sum 
  $I_1^N+I_2^{N}(t)+I_3^{N}(t)$ in Proposition~1.
  The  only difference is that $I_1^N$ tends to the integral along~${\R}^3$ 
   of the density of three  independent standard Gaussians, which
equals~$1$. 
   Thus, in fact, 
   $\bar{Z}_N(t, \widetilde{\epsilon})$ converges to~$1$
   uniformly in $[T_0, T]$ and \eqv(3exp) is obvious.
  Hence, for  all $t\in[T_0, T]$
$$
\frac{ \E \bar{Z}^2_N(t,\widetilde{\epsilon} ) }
{\big(\E  
 \bar{Z}_N(t,\widetilde{\epsilon})\big)^2}\leq           
\frac{\big(\E\bar{Z}^3_N(t,\widetilde{\epsilon}) \big)^{2/3}}
{\big(\E  
 \bar{Z}_N(t,\widetilde{\epsilon})\big)^2}\leq 
\frac{K_2^{2/3}}{K_1^2}:=K_3.      
\Eq(2exp)
  $$
 Then  starting from the Paley-Zygmund  inequality and 
  finally applying the concentration of measure 
  inequality \eqv(conc) with $u= 
Nt/2-\E \ln \E_{\s}e^{H_N(t, \s)}+\ln(K_1/2)$, we deduce
$$
\eqalign{
1/4K_3&\leq  \frac{\E\bar{Z}^2_N(t, \widetilde{\epsilon})}
{4\big(\E\bar{Z}_N(t, \widetilde{\epsilon})\big)^2}
\leq \P\{ \bar{Z}_N(t, \widetilde{\epsilon}) > 
\E\bar{Z}_N(t, \widetilde{\epsilon})/2\}
\leq
\vphantom{\sum}\P\{ \E_{\s} e^{H_N(t, \s)} > K_1 e^{Nt/2}/2 \}\cr
\vphantom{\sum}&= \P\{ \ln\E_{\s} e^{H_N(t, \s)}-\E \ln \E_{\s}e^{H_N(t,
\s)}>
Nt/2-\E \ln \E_{\s}e^{H_N(t, \s)}+\ln(K_1/2)\}\cr
&\leq  \exp\{ 
[Nt/2-\E \ln \E_{\s}e^{H_N(t, \s)}+\ln(K_1/2)]^2/2Nt\},
}
$$
 from where \eqv(est) follows.
    Finally,~\eqv(LK1), \eqv(IL) and \eqv(est)    
  together imply
$$
\eqalign{
\E
&\frac{\E_{\s} e^{H_N(t, \s)} \1_{\{H_N(t,\s)>Nt(1+\epsilon)\}} }
{ \E_{\s}e^{H_N(t, \s)} } \cr
\vphantom{\sum}\leq &
e^{Nt\epsilon^2/4+K\sqrt{N}}
\E\E_{\s}e^{H_N(t, \s)-Nt/2}\1_{\{H_N(t,\s)>Nt(1+\epsilon)\}}
+ e^{-Nt\epsilon^4/32}\cr
\leq & e^{-Nt^2/4+K\sqrt{N}}+e^{-Nt\epsilon^4/32},
}
\Eq(LK4)
$$     
  and the proposition is proved.\endproof

\noindent{\bf Proof of \eqv(1.21).}
  To complete the proof of Theorem 1.3, it remains to show that 
$$
  \lim_{p\uparrow +\infty}\inf_{m_1, m_2, m_3\in {\cal A}} Y_{p}(m_1, m_2, m_3)= 2\ln 2.
\Eq(ppp)
$$ 
 After elaborating the functions $S_p(m_1,m_2,m_3)$ and $R_p(m_1, m_2, m_3)$, we get:
$$
\eqalign{ 
U_p(m_1, m_2, m_3)=& I(m_1, m_2, m_3)(1+m_3^p)\cr
&\times{}\Big[ \Big(
4\Big(1+m_3^p+\frac{m_1^p+m_2^p}{2}\Big)^2+\frac{(m_1^p-m_2^p)^2
(1+m_3^p)}{(1-m_3^p)}\Big)^{1/2}\cr
&\qquad\qquad\qquad\qquad\qquad
{}-\frac{(m_1^p-m_2^p)^2}{2(1-m_3^{p})}
-m_1^p m_2^p-(2+m_3^p)\Big]^{-1}.
}
\Eq(uu)  
$$  
   It follows from~\eqv(uu)  that
for any $p=2k>2$ and any sequence 
 $(m_{1,n}, m_{2, n}, m_{3, n})\in{\cal A}$ 
 such that $m_{1, n}\rightarrow 1$, $m_{2, n}\rightarrow 1$, 
  $m_{3, n}\rightarrow 1$, as $n\rightarrow \infty$,
$$
\lim_{n\uparrow + \infty}
  Y_p(m_{1,n}, m_{2,n}, m_{3,n})=2\ln 2.
\Eq(sup)  
$$
(In fact, by the definition of~${\cal A}$ we have     
   $||m_1|-|m_2||\leq 1-|m_3|$  for all 
    $(m_1, m_2, m_3)\in {\cal A}$, whence
  $(m_{1,n}^p-m_{2,n}^p)^2=o(1-m_{3,n}^p)$.)
 Thus 
$$\lim \sup_{p\uparrow + \infty} \inf_{m_1, m_2, m_3\in A}Y_p(m_1, m_2, m_3)\leq 2\ln 2.$$  
    This fact and the next Proposition~\thv(prp4) together
  imply~\eqv(ppp). \endproof
 
\proposition{\TH(prp4)}{\it
 Let $\{p_n\}$ be a sequence of positive even numbers,
   $p_{n}\uparrow + \infty$.
  Assume that the sequence $(m_{1, n}, m_{2, n}, m_{3, n})\in {\cal A}$ 
  satisfies one of the following conditions:
\item{(i)} $|m_{1, n}|\rightarrow 1$,
     $|m_{2, n}|\rightarrow 1$, $|m_{3, n}|\rightarrow 1$;
\item{(ii)}  there exist $\delta>0$ and a pair $i$ and $j$,
$i,j=1,2,3$,
   $i\ne j$, such that 
    $|m_{i, n}|\rightarrow 1$ and $|m_{j, n}|\leq 1-\delta$ 
   for all sufficiently large~$n$;
\item{(iii)} there exists $\delta>0$ such that 
     $|m_{1, n}|\leq 1-\delta$, 
    $|m_{2, n}|\leq 1-\delta$, 
   $|m_{3, n}|\leq 1-\delta$  for all sufficiently large~$n$.        
   Then 
$$
\lim \inf_{n\uparrow + \infty} Y_{p_n}(m_{1, n}, m_{2, n}, 
  m_{3, n})\geq 2\ln 2.
\Eq(sp)  
$$      
}

\noindent{\bf Proof:} In the cases (i) and (iii)  it suffices 
  to substitute the sequence $(m_{1,n}, m_{2,n},m_{3,n})$
   into the function $I(m_1, m_2, m_3)(2/3+(m_1^p+m_2^p+m_3^p)^{-1})$.
In  case (ii) assume that e.\ g.\ $|m_{3, n}|\rightarrow 1$ 
and $|m_{1, n}|\leq 1-\delta$.
   Then 
$m_{1,n}^{p_n}=o(1)$.
 By definition of the set ${\cal A}$ we obtain 
 $ ||m_{1, n}|-|m_{2, n}||\leq 1-|m_{3, n}|\rightarrow 0 $ 
   as $n\uparrow + \infty$, thus
$m_{2,n}^{p_n}=o(1)$ and
  $(m_{1,n}^{p_n}-m_{2,n}^{p_n})^2/(1-m_{3,n}^{p_n})=o(1)$.
    Moreover,
 if $m_{3,n}\rightarrow 1$, then $m_{1,n}-m_{2,n}\rightarrow 0$ and
      if $m_{3,n}\rightarrow -1$, then $m_{1,n}+m_{2,n}\rightarrow 0$
  and therefore in both of these cases
$
 \lim \inf_{n\uparrow + \infty} I(m_{1, n}, m_{2, n}, m_{3,
n})\geq \ln 2.$
    This  yields 
$$ \eqalign{
&\lim\inf_{n \uparrow + \infty} Y_{p_n}(m_{1, n}, m_{2, n}, m_{3, n})\cr
&\quad \geq
    \lim\inf_{n\uparrow + \infty} U_{p_n}(m_{1,n}, m_{2,n}, m_{3,n})\geq
    \lim\inf_{n \uparrow + \infty} \ln 2 \frac{1+m_{3,n}^{p_n}}{m^{p_n}_{3,
n}+o(1)}
\geq 2\ln 2
}$$ 
and the proposition is proved. \endproof

\bigskip

\chap{3. The fluctuations of the partition function in the REM.}3

Amazingly enough, the simplest of all our models, the REM, will 
be seen to offer in some sense the most interesting behaviour 
with regard to the fluctuations of the free energy. The main surprise here 
will be the existence of an intermediate region of temperatures where 
a CLT does not hold, but there a non-standard limit theorem will be proven.

We begin with the proof of (i) of Theorem 1.4.

\proposition {\TH(REM.1)}{\it  Whenever $0\leq \b<\sqrt{\ln 2/2}$, 
$$
e^{\frac N2(\ln 2-\b^2)}  \ln \frac{ Z_{\beta,N}}{ \E Z_{\beta,N} } \limlaw
\NN(0,1).
\Eq(R.1)
$$
}

\noindent{\bf Proof.} 
This result will follow from the standard CLT for triangular arrays.
Let us first write 
$$
\ln \frac{Z_{\beta,N}}{\E Z_{\beta,N}}=\ln\Bigl(1+\frac{Z_{\beta,N}-\E Z_{\b,N}}{\E Z_{\b,N}}\Bigr).
\Eq(R.2)
$$
We will show that the second term in the logarithm 
properly normalized will converge to a normal random variable. To see this,
write
$$
\frac{Z_{\beta,N}-\E Z_{\beta,N}}{\E Z_{\b,N}} =
\sum_{\s\in \SS_N} e^{-N(\ln 2+\b^2/2) } 
\left(e^{\b\sqrt N X_\s}-e^{N\b^2/2}\right)
\equiv \sum_{\s\in \SS_N} \YY_N(\s).
\Eq(R.3)
$$
Note that $\E \YY_N(\s) =0$ and 
$
\E \YY^2_{N}(\s) = e^{-N (2\ln 2 -\b^2)}[1-e^{-N\b^2}]
$
and thus 
$$
\E\Bigl( \frac{Z_{\b,N}-\E Z_{\b,N}}{\E Z_{\b,N}}\Bigr)^2= e^{-N (\ln 2
-\b^2)}[1-e^{-N\b^2}].
\Eq(R.5)
$$
Therefore we can 
write 
$$
 \frac{Z_{\b,N}-\E Z_{\b,N}}{\E Z_{\b,N}}
=e^{-\frac N2(\ln 2-\b^2)}\sqrt{1-e^{-N\b^2}}
\frac 1{2^{N/2}}\sum_{\s\in\SS_N} \wt \YY_N(\s),
\Eq(R.6)
$$
where $\wt \YY_N(\s)= e^{\frac{N}{2}(2\ln 2 -\b^2)}[1-e^{-N\b^2}]^{-1/2}
\YY_N(\s)$
has mean zero and variance one. By the CLT for triangular arrays (see [Shi]),
it follows readily that 
$$
\frac 1{2^{N/2}}\sum_{\s\in\SS_N} \wt \YY_N(\s)
\limlaw \NN(0,1)
\Eq(R.7)
$$
if the Lindeberg condition holds, that is in this case if for any $\e>0$,
$$
\lim_{N\uparrow + 0}\E \wt \YY_{N}^{2}(\s)\1_{\{|\wt{\YY}_N(\s)|\geq \e 2^{N/2}\}}=0.
\Eq(R.8)
$$
But 
$$
\eqalign{
\E \wt \YY^{2}_{N}(\s)\1_{\{|\wt{\YY}_N(\s)|\geq \e 2^{N/2}\}}
&=\frac 1{\sqrt {2\pi}(1-e^{-N\b^2})}
e^{-2N\b^2}\!\!\!\!\!\!\!\int\limits_{\sqrt N(\frac{\ln 2}{2\b}+\b)+\frac{\ln \e}{\sqrt N\b}+o(\frac{1}{\sqrt{N}})}^{\infty} 
\!\!\!\!\!\!\!e^{2\sqrt N\b z -\frac {z^2}2} { d}z+o(1)\cr
&=\frac 1{\sqrt {2\pi} (1-e^{-N\b^2})}\!\!\!\!\!\!\!
\int\limits_{\sqrt N(\frac{\ln 2}{2\b}-\b)+\frac{\ln \e}{\sqrt N\b}+o(\frac{1}{\sqrt{N}})}^{\infty} 
\!\!\!\!\!\!\! e^{-\frac {z^2}2}  {d}z+o(1).\cr
}
\Eq(R.9)
$$
It is easy to check that the latter integral converges to zero if and 
only if $\b^2<{\ln 2}/2$.
  Using now the fact that $e^{x}=1+x+o(x)$ as $x\rightarrow 0$,
 it is now a trivial matter to deduce the assertion of the 
  proposition. \endproof

Since the Lindeberg condition clearly fails for $2 \b^2\geq \ln 2$, it is
clear
 that we cannot expect a simple CLT beyond this regime. Such a failure of 
a CLT is always a problem related to ``heavy tails'', and results from the
fact
that extremal events begin to influence the fluctuations of the sum. It
appears
therefore reasonable to separate form the sum the terms where $X_\s$ is 
anomalously large. For Gaussian r.v.'s it is well known that the right scale
of separation is given by $u_N(x)$ defined
by
$$
2^N \int\limits_{u_N(x)}^\infty \frac {dz}{\sqrt{2\pi}} e^{-z^2/2} =e^{-x}
\Eq(R.11)
$$
which (for $x>-\ln N/\ln 2$) is equal to (see e.g. [LLR])
$$
u_N(x)= \sqrt{2N\ln 2} +\frac{x}{\sqrt {2N\ln 2}}
-\frac{\ln (N\ln 2)+\ln 4\pi}{2\sqrt{2N\ln 2}}+o(1/\sqrt N),
\Eq(R.12)
$$
$x\in\R$ is a parameter. Let us now define 
$$
Z^x_{N,\b}\equiv \E_\s e^{\b \sqrt N X_\s}\1_{\{X_\s\leq u_N(x)\}}.
\Eq(R.13)
$$
We may write
$$
\frac {Z_{\b,N}}{\E Z_{\b,N}}=1+\frac{\Z_{\b,N}^x-\E Z_{\b,N}^x}{\E Z_{\b,N}}
+\frac {Z_{\b,N}-Z_{\b,N}^x -\E (Z_{\b,N}-Z_{\b,N}^x)}{\E Z_{\b,N}}
\Eq(R.14)
$$
Let us first consider the last summand. We introduce the random variable
$$
\WW_N(x)=\frac {Z_{\b,N}-Z_{\b,N}^x}{\E Z_{\b,N}} = e^{-N(\ln 2+\b^2/2)}
\sum_{\s\in\SS_N} e^{\b\sqrt N X_\s} \1_{\{X_\s>u_N(x)\}}
\Eq(R.15)
$$
It will be convenient to rewrite this as (we ignore the subleading 
corrections to $u_N(x)$ and only keep the explicit representation \eqv(R.12))
$$
\eqalign{
\WW_N(x)&= e^{-N(\ln 2+\b^2/2)}\sum_{\s\in\SS_N} 
e^{\b\sqrt N u_N(u^{-1}_N( X_\s))} \1_{\{u_N^{-1}(X_\s)>x\}}\cr
&=e^{-N(\ln 2+\b^2/2)} e^{\b N\sqrt{2\ln 2} -
\b \frac{\ln (N\ln 2)+\ln 4\pi}{2\sqrt{2\ln 2}}}
\sum_{\s\in\SS_N} e^{\frac {\b}{\sqrt{2\ln 2}} u_N^{-1}(X_\s)}
\1_{\{u_N^{-1}(X_\s)>x\}}.
}
\Eq(R.16)
$$
Let us now introduce the point process on $\R$ given by 
$$
\PP_N \equiv \sum_{\s\in \SS_N} \d_{u_N^{-1}(X_\s)}.
\Eq(R.17)
$$
A classical result from the theory of extreme order statistics (see e.g.
[LLR]) asserts that
the point process $\PP_N$ converges weakly to a Poisson point process on 
$\R$ with intensity measure $e^{-x}dx$. We can, of course, write 
$$
\sum_{\s\in\SS_N} e^{\frac {\b}{\sqrt{2\ln 2}} u_N^{-1}(X_\s)}
\1_{\{u_N^{-1}(X_\s)>x\}}
=\int\limits_{x}^\infty e^{\a z}\PP_N(dz),
\Eq(R.18)
$$
where we set $\a\equiv\b/\sqrt {2\ln 2}$.
Clearly, the weak convergence of $\PP_N$ to $\PP$ implies 
convergence in law of the right hand side of \eqv(R.18), provided
that $e^{\a x} $ is integrable on $[x,\infty)$ w.r.t.\ the Poisson process
with intensity $e^{-x}$. 
This is, in fact never a problem: the Poisson point process has almost
surely 
support on a finite set, and therefore $e^{\a x}$ always a.s.\ integrable. 
Note, however, that for $\b\geq \sqrt{ 2\ln 2}$ the mean of the integral is 
infinite, indicating the passage to the low temperature regime. Note also
that 
the variance of the integral is finite exactly if $\a<1/2$, i.e. 
$\b^2<\ln 2/2$, i.e. when the CLT holds. On the
other
hand, the mean of the integral diverges if $x\downarrow\-\infty$; note 
that
at minus infinity the points of the Poisson point process accumulate, and
there
is no finite support argument as before that would assure the
existence if $x$ is taken to $-\infty$. The following lemma provides the first
step in the proof of part  (ii) of Theorem 1.4 and of Theorem 1.5:

\lemma{\TH(REM.2)} {\it Let $\WW_N(x),\a$ be defined above, and let $\PP$ be the 
Poisson point process with intensity measure $e^{-z} dz$. Then 
$$
e^{\frac N2 (\sqrt{2\ln 2}-\b)^2  +
 \frac\a 2[\ln (N\ln 2)+\ln 4\pi]} \WW_N(x)
\limlaw \int\limits_{x}^\infty e^{\a z}\PP(dz).
\Eq(R.19)
$$
}

\remark Note that the mean of the right hand side is finite if and
only of $\b<\sqrt {2\ln 2}$. Thus only in that case does this lemma
also allow to deal with the centered variable appearing in \eqv(R.14).

We now need to turn to the remaining term,
$$
\frac {Z_{\b,N}^x-\E Z_{\b,N}^x}{\E Z_{\b,N}}=\frac{\VV_N(x)}{\E Z_{\b,N}},
\Eq(R.20bis)
$$
where
$$
\VV_N(x)\equiv {Z^x_{\b,N}-\E Z^x_{\b,N}}.
\Eq(R.20)
$$
One might first hope that this term upon proper scaling would converge to a 
Gaussian; however, one can easily check that this is not the case (the 
Lindeberg condition will not be verified). However, it will not be hard to 
compute all moments of this term:

\lemma{\TH(REM.3)}
{\it Let $\VV_N(x)$ be defined by \eqv(R.20). Then for $\a>1/2$
and any  integer $k\geq 2$
$$
\eqalign{\lim_{N\uparrow +\infty}
\frac{ 
\E [\VV_N(x)]^k}{\left[2^{-N}e^{N\beta\sqrt{2\ln 2}-
\frac\a 2[\ln (N\ln 2)+\ln 4\pi]}
\right]^k}
 &=\sum_{i=1}^{k}\frac{1}{i!}\sum_{{\ell_1\geq 2,\dots,\ell_i\geq 2}
\atop{\sum_{j}\ell_j=k}}  \frac{k!}{\ell_1!\dots\ell_i!}
\frac{ e^{(k\a-i) x}}
{(\ell_1\a-1)\dots(\ell_i\a-1)}.\cr
}
\Eq(R.21)
$$
   For $\alpha=1/2$,  we have for $k$ even 
$$
\eqalign{ \lim_{N\uparrow +\infty}
\frac{ 
\E [\VV_N(x)]^k}{\left[2^{-N}e^{N\beta\sqrt{2\ln 2}}
\right]^k}
 &=\frac{k!}{(k/2)!\, 2^{k}} =\frac{(k-1)!!}{2^{k/2}}
}
\Eq(R.21bis)
$$
  and for $k$ odd
$$
 \lim_{N\uparrow +\infty}
\frac{ 
\E [\VV_N(x)]^k}{\left[ 2^{-N} e^{N\beta\sqrt{2\ln 2}}
\right]^k}
 =0
\Eq(R.21bis') 
$$  
(which are the moments of the normal distribution with variance $1/2$).
}

{\it Proof.} This is a pure computation. 
Set $T_N(\s)\equiv e^{\b \sqrt N X_\s}\1_{\{X_\s\leq u_N(x)\}}$. Note that 
for $\b<\sqrt{2\ln 2}$
$$
\E T_N(\s)=\int\limits_{-\infty}^{u_N(x)}\frac {dz}{\sqrt 2\pi} e^{-\frac {z^2}2+
\b\sqrt N z}
=e^{N\b^2/2}\Bigl(1-\int\limits_{u_N(x)-\b \sqrt N}^\infty  \frac {dz}{\sqrt 2\pi}
e^{-\frac {z^2}2}\Bigr)
\sim e^{\beta^2 N/2}.
\Eq(R.22)
$$
while for $\b>\sqrt{2\ln 2}$ and all $k\geq 1$, and for 
$\b>\sqrt{\ln 2/2}$ and for $k\geq 2$,
$$
\eqalign{
&\E [T_N(\s)]^k=\int\limits_{-\infty}^{u_N(x)} \frac {dz}{\sqrt 2\pi} e^{-\frac
{z^2}2+
k\b\sqrt N z}
=e^{Nk^2\b^2/2}\int\limits_{-\infty}^{u_N(x)-k\b \sqrt N}  \frac {dz}{\sqrt 2\pi} 
e^{-\frac {z^2}2}\cr
&\quad\sim e^{Nk^2\b^2/2} \frac {e^{-(u_N(x)-k\b \sqrt N)^2/2}}{\sqrt{2\pi}(k\b\sqrt N-u_N(x))}
\sim \frac {2^{-N} e^{-x}}{k\a-1}e^{k[\b\sqrt {2\ln 2} N+\a x-
\frac \a 2[\ln (N\ln 2)+\ln 4\pi]]}.
}
\Eq(R.23)
$$
  Formula \eqv(R.23) is also valid for $\beta=\sqrt{2\ln 2}$ with $k>1$
 and  for $\beta=\sqrt{\ln 2/2}$ with $k>2$.
 It is easy to see from the computations above that for $\beta=\sqrt{2\ln
2}$ with $k=1$ 
 and also for $\beta=\sqrt{\ln 2/2}$ with $k=2$ we have 
$$\E [T_N(\s)]^{k} \sim \frac{e^{k^2\beta^2 N/2}}{2}=
\frac{2^{-N} e^{-x}}{2}e^{k[\beta\sqrt{2\ln 2}N+\alpha x]}.
\Eq(R.23bis)
$$
We set 
$\wt T_N(\s)\equiv 2^{-N} T_N(\s)$; by \eqv(R.23) we get for
  $\beta>\sqrt{\ln 2/2}$ with $k\geq 2$  and also for
$\beta>\sqrt{2\ln 2}$ with $k\geq 1$
$$ \E [\wt{T}_N(\s)]^{k}=
\frac {2^{-N} e^{-x}}{k\a-1}e^{k[\b\sqrt {2\ln 2} N-\ln 2+\a x-
\frac \a 2[\ln (N\ln 2)+\ln 4\pi]]}.
\Eq(R.24)
$$
   This formula is also true  for $\beta=\sqrt{\ln 2/2}$, $k>2$ and
$\beta=\sqrt{2\ln 2}$, $k>1$.
 For $\beta=\sqrt{2\ln 2}$ and $k=1$ and also for $\beta=\sqrt{\ln
2/2}$ and 
    $k=2$  by \eqv(R.23bis)
 $$ \E [\wt{T}_N(\s)]^{k}=
\frac {2^{-N} e^{-x}}{2}e^{k[\b\sqrt {2\ln 2} N-\ln 2+\a x]}.
\Eq(R.24bis)
$$ 
Now 
$$\eqalign{
\E\left[\VV_N(x)\right]^k&=
\E\Bigl(\sum_{\s\in\SS_N}[\wt T_N(\s)-\E\wt T_N(\s)]
\Bigr)^k
=\sum_{\s_1,\dots,\s_k\in \SS_N}
\E\prod_{i=1}^k\bigl[\wt T_N(\s_i)-\E\wt T_N(\s_i)\bigr]\cr
&=
\sum_{i=1}^k \sum_{{\ell_1,\dots,\ell_i\geq 2}\atop{\sum_{j}\ell_j=k}}
\frac{k!}{\ell_1!\dots\ell_i!} {2^{N}\choose i}
\E [\wt T_N(\s)-\E\wt T_N(\s)]^{\ell_1}\dots 
\E [\wt T_N(\s)-\E\wt T_N(\s)]^{\ell_i}.
}
\Eq(R.25)
$$

Note finally that for $l\geq 2$ and $\beta\geq \sqrt{\ln 2/2}$
$$
\eqalign{
\E\bigl[\wt T_N(\s)-\E\wt T_N(\s)\bigr]^\ell
&=\sum_{j=1}^\ell (-1)^{j} {\ell\choose j} \E \wt T_N(\s)^{\ell-j} 
[\E  \wt T_N(\s)]^j
\sim  \E \wt T_N(\s)^{\ell}. 
}
\Eq(R.26)
$$
 In fact, if $\sqrt{\ln 2/2}\leq \beta < \sqrt{2\ln 2}$, $l\geq 2$, $j\geq
1$,
  $j\ne l-1, l$, 
  then by \eqv(R.22) and \eqv(R.24), \eqv(R.24bis)
$$ \frac{ \E[T^{l-j}_N(\s)][ \E T_N(\s)]^j}{\E[T^{l}_N(\s)]}
    =
   =e^{Nj(\beta^2/2-\beta\sqrt{2\ln 2})}) O\left(N^{\a j/2}\right)
\Eq(R.26a)
$$
  For $l\geq 2$, $j=l-1,l$
$$ 
\eqalign{
&\frac{ \E[T^{l-j}_N(\s)][ \E T_N(\s)]^j}{\E[T^{l}_N(\s)]}
 e^{ Nl(\beta^2/2-\beta \sqrt{2\ln 2})+N\ln 2}
O\left(N^{\a l/2}\right)\leq e^{-N\ln 2/2}) N^{\a} 
}
\Eq(R.26b)
$$
 For $\beta\geq \sqrt{2\ln 2}$, $l\geq 2$ and $j\geq 1$  by \eqv(R.24) and
\eqv(R.24bis)
$$ 
\frac{ \E[T^{l-j}_N(\s)][ \E T_N(\s)]^j }{ \E[T^{l}_N(\s)] }
    =O(2^{-Nj}).
\Eq(R.26c)
$$
Thus  for $l\geq 2$ and $\beta>\sqrt{\ln 2/2}$
  and also for $l\geq 3$ and $\beta=\sqrt{\ln 2/2}$
$$
\E \bigl[\wt T_N(\s)-\E\wt T_N(\s)\bigr]^\ell
= \frac {2^{-N} e^{-x}}{k\a-1} \bigl[
 2^{-N} e^{N\b \sqrt {2\ln 2}}e^{\a x}e^{-\frac \a 2[\ln (N\ln 2)+\ln 4\pi]}
\bigr]^{\ell}.
\Eq(R.27)
$$
 Inserting this result into \eqv(R.25) gives the assertion of the 
lemma \eqv(R.21).
  
 For $\beta=\sqrt{\ln 2/2}$ and
 $l=2$ by \eqv(R.24bis) we have
$$
\E\bigl[\wt T_N(\s)-\E\wt T_N(\s)\bigr]^2
= \frac {2^{-N} e^{-x}}{2}\bigl[
2^{-N} e^{N\b \sqrt {2\ln 2}}e^{\a x}
\bigr]^{2}.
\Eq(R.27bis)
$$   
  Inserting this formula into \eqv(R.25) we see, that 
  the term with $l_1, \ldots, l_i=2$, $i=k/2$ brings the main contribution
  to the sum, 
  and all others are of smaller order, because 
  of the polynomial terms $e^{-l\frac{\alpha}{2}\ln (N \ln 2)}$
 in \eqv(R.27). 
 This implies~\eqv(R.21bis) and \eqv(R.21bis') and the lemma is proved.
\endproof

\remark One sees that if we let $x\downarrow -\infty$, and rescale 
properly, the corresponding moments converge to that of a centered Gaussian 
r.v. This could alternatively be seen by checking that the Lindeberg 
condition holds for the truncated variables provided $x\leq -2\ln\ln 2^N$. 

A standard consequence of Lemma \thv(REM.3) is the weak convergence of
the normalized version of $\VV_N(x)$:

\corollary{\TH(REM.4)}{\it For $\sqrt{\ln 2/2}<\b$,
$$
e^{\frac N2(\sqrt{2\ln 2}-\b)^2+\frac\a 2[\ln (N\ln 2)+\ln 4\pi]}
\frac{\VV_N(x)}{\E Z_{\b,N}}\limlaw \VV(x,\a),
\Eq(R.28)
$$
where $\VV(x,\a)$ is the random variable with mean zero and $k$th moments
given by the right hand side of \eqv(R.21).
  For $\beta=\sqrt{\ln 2/2}$
$$
\sqrt{2} e^{\frac N2(\sqrt{2\ln 2}-\b)^2}
\frac{\VV_N(x)}{\E Z_{\b,N}}\limlaw \NN(0,1).
\Eq(R.28bis)
$$
}
The next proposition will imply (ii) of Theorem~1.4.

\proposition {\TH(REM.5)}{\it Let $\sqrt{\ln 2/2}<\b<\sqrt{2\ln 2}$. 
Then  for $x\in \R$ chosen arbitrarily,
$$
e^{\frac N2(\sqrt{2\ln 2}-\b)^2+\frac\a 2[\ln (N\ln 2)+\ln 4\pi]} \ln 
\frac {Z_{\beta, N}}{\E Z_{\beta,N}} \limlaw \VV(x,\a) +\int\limits_{x}^\infty
e^{\a z}\PP(dz) - \int\limits_{x}^\infty
e^{\a z}e^{-z}dz,  
\Eq(R.29)
$$
where $\VV(x,\a)$ and $\PP$ are independent random variables.
}

\noindent{\bf Proof.}
 \eqv(R.29) would be immediate from Lemma \thv(REM.2) and Corollary
\thv(REM.4), if $\WW_N(x)$ and $\VV_N(x)$ were independent. However, while this 
is not true, they are not far from independent. To see this, note that 
if we condition on the number of variables $X_\s$, $n_N(x)$,  that 
exceed $u_N(x)$, 
the decomposition in \eqv(R.14) is independent. On the other hand, 
one readily verifies that Corollary \eqv(REM.4) 
also holds under the conditional law $\P[\cdot|n_N(x)=n]$,
 for any finite $n$, with the same right hand side $\VV(x,\a)$. But this implies
that the limit can be written as the sum of two independent random variables,
as desired.
\endproof

Since for $\b^2>\ln 2/2$, $\a >1/2$, one sees that $\E \VV(x,\a)^2=
e^{x(2\a-1)}  /(2\a-1)$ tends to zero as $x\downarrow -\infty$. Therefore 
we see that 
$$
\VV(x,\a) =_\DD \lim_{y\uparrow + \infty}
\int\limits_{-y}^x
e^{\a z}\PP(dz) - \int\limits_{-y}^x
e^{\a z}e^{-z}dz  
\Eq(R.30)
$$
which means that we can give sense to the Poisson integral
$
\int_{-\infty}^\infty e^{\a z}(\PP(dz)-e^{-z}dz)
$
We see that Propositions \thv(REM.1) and \thv(REM.5) imply Theorem 1.4. 
\endproof\endproof


\remark The appearance of the intermediate region with non-Gaussian 
fluctuations  may appear surprising in view of the fact that in the $p$-spin
models, we could prove the CLT up to a much higher value of $\b$, in fact 
up to almost the critical value. The reason, however, lies in the fact that
in the $p$-spin model the Gaussian part of the fluctuation is always 
on a polynomial scale in $N$, while the truncation error ($(Z_{\b,N}-Z_{\b,N}^T)/\E
Z_{\b,N}$)
is exponentially small even when we truncate at $\b(1+\e) \sqrt N$,
way below where we truncate in the REM. This means that the CLT contribution
will always dominate the extremal fluctuations. In the REM everything is 
exponentially small, and while a sufficiently truncated partition function 
gives a Gaussian contribution, this is dominated by the larger extremal 
fluctuations in the intermediate regime. In other words, the extra
correlations
in the $p$-spin models strengthen the Gaussian fluctuations more than the 
extremal ones which sounds intuitive.



We now turn to the 

\noindent{\bf Proof of  {Theorem 1.5}}.
We will see that the computions above almost suffice to 
conclude the low temperature case as well.
 With the notations from above, we write
$$
Z_{\beta,N}=Z_{\b,N}^x+(Z_{\b,N}-Z_{\b,N}^x)
\Eq(R.35)
$$
Clearly for $\beta \geq \sqrt{2\ln 2}$ 
$$
Z_{\b,N}-Z_{\b,N}^x= 
e^{ N\left[\b \sqrt{2\ln 2}-\ln 2\right]-\frac{\a}2[\ln (N\ln 2)+\ln 4\pi]}
\sum_{\s\in\SS_N}\1_{\{u_N^{-1}(\s)>x\}} e^{\a u_N^{-1}(X_\s)}
\Eq(R.36)
$$
so that for any $x\in \R$,
$$
(Z_{\b,N}-Z_{\b,N}^x)
e^{ -N\left[\b \sqrt{2\ln 2}-\ln 2\right]+\frac{\a}2[\ln (N\ln 2)+\ln
4\pi]}\limlaw \int\limits_{x}^\infty e^{\a z} \PP(dz).
\Eq(R.37)
$$
Now write 
$$
Z_{\b,N}^x=\E Z_{\b,N}^x\Bigl(1+\frac {Z_{\b,N}^x-\E Z_{\b,N}^x}{\E Z_{\b,N}^x}\Bigr).
\Eq(R.38)
$$
  Let us first treat the case $\beta>\sqrt{2\ln 2}$.
By~\eqv(R.23)  we have
$$
\E Z_{\b,N}^x 
\sim \frac {2^{-N} e^{-x}}{\a-1}e^{\b\sqrt {2\ln 2} N+\a x-
\frac \a 2[\ln (N\ln 2)+\ln 4\pi]}.
\Eq(R.39)
$$
 Thus 
$$
e^{-N\left[\b \sqrt{2\ln 2}-\ln 2\right]+\frac{\a}{2}[\ln (N\ln 2)+\ln
4\pi]} Z_{\b,N}^x
=\frac{e^{x(\a-1)}}{\a-1}\Big(1+ \frac{Z_{\b,N}^x-\E Z_{\b,N}^x}{\E Z_{\b,N}^x}\Big)(1+o(1)).
\Eq(R.40)
$$
Using Lemma \thv(REM.3) we see that now $\frac{Z_{\b,N}^x-\E Z_{\b,N}^x}{\E
Z_{\b,N}^x}\frac{e^{x(\alpha-1)}}{\alpha-1}$
converges in distribution to a random variable with moments given by
the right hand side of~\eqv(R.21). Moreover, as $x\downarrow -\infty$,
this variable converges to zero in probability.
Since the same is true for the prefactor, the assertion of the theorem is 
now immediate.

  Let us consider now the case $\beta=\sqrt{2\ln 2}$.
 Proceeding as in \eqv(R.23),
 $$
\E Z_{\beta,N}^0
= \frac{2^{N}}{\sqrt{2\pi}} 
\int\limits_{-\infty}^{u_N(0)-\sqrt{2N\ln2}}e^{-z^2/2}dz
= 2^N\Big(\frac12 - \frac {\ln (N\ln 2)+\ln 4\pi}{4\sqrt{N\pi\ln2}}+O
\Big(\frac{(\ln N)^2}{N}\Big)\Big).
\Eq(R.41)
$$
  We use the decomposition
$$ 
Z_{\beta, N}=Z_{\b,N}-Z_{\b,N}^0+ \E Z_{\b,N}^0+(Z_{\b,N}^0-\E Z_{\b,N}^0).
\Eq(R.42)
$$  
By  \eqv(R.41), $\E Z^0_{\b,N}/\E Z_{\b,N}\sim 1/2$. By \eqv(R.16), we see easily 
that 
$$
 \frac{Z_{\b,N}-Z_{\b,N}^0}{\E Z_{\b,N}}=\WW_N(x)\rightarrow 0 \text {a.s.} 
\Eq(R.43)
$$      
even though $\E \WW_N(0)=1/2$! Thus the more precise statement
consists in saying that
$$
e^{\frac 12[\ln(N\ln 2)+\ln 4\pi]} \WW_N(0)\limlaw \int\limits_{0}^\infty
 e^z\PP(dz).
\Eq(R.44)
$$
Note that of course the limiting varaible has infinite mean, but is
a.s. finite.
Finally, by Corollary \thv(REM.4), 
$$
e^{\frac 12[\ln(N\ln 2)+\ln 4\pi]}\frac{Z_{\b,N}^0-\E Z_{\b,N}^0}{\E Z_{\b,N}} \limlaw \VV(0,1)
\Eq(R.45)
$$
The same arguments as those given after Proposition \thv(REM.5) allow us to 
identify $\VV(0,1)$ with the Centered Poisson integral
$
\int_{-\infty}^0 e^{z}\left(\PP(dz)-e^{-z}dz\right).
$
This implies  \eqv(1.25). \eqv(1.25a) is an immediate corollary.
This concludes the proof of Theorem 1.5.\endproof\endproof

\bigskip

\chap{Appendix 1. Some remarks on the case $p$ odd}4

\noindent{\bf Conjecture \TH(cj)}{\it
Let $p=2k+1$, $k\geq 1$. There exists $\beta_p>0$ such that 
  for all $\beta<\beta_p$
$$
N^{(p-2)/2}\ln\frac{Z_{\beta,N}}{\E Z_{\beta, N}}
\rightarrow
     M_{\infty}(\sqrt{\beta})
\Eq(podd) 
 $$
 in distribution as $N\uparrow +\infty$,   
where $M_{\infty}(t)$ is a centred Gaussian process
 with independent increments and
  $$ \E(M_{\infty}(t)-M_{\infty}(s))^{2}=\frac{(t^2-s^2)[(2p-1)!!]}{2},$$
  Moreover
$
\beta_p \rightarrow \sqrt{2\ln 2},$ as $p\uparrow + \infty$. 
}

\noindent {\bf Discussion.} 
   Let us try to adapt the martingale method in this case.
  This leads to 
$$\eqalign{
 V_N(t)=&\E_{\s, \s'} \Big(N \Big(R_N(\s,\s')\Big)^p -
         N^{2-p} t \E\xi^{2p}\Big)e^{H_N(t,\s)+H_N(t,\s')-Nt}.
}
$$
  Then
$$
N^{p-2} \E V_N(t) =
\sum_{m=0,\pm 1/N, \ldots, \pm 1} 
 \Big(N^{p-1} m^p- t\E\xi^{2p}\Big)e^{tN m^p}
                               \P(\s\cdot\s'=mN).
\Eq(d1) 
 $$
 It is easy to show that $N^{p-2} \E V_N(t) \rightarrow 0$ 
 as $N \uparrow + \infty$ for all $t$ such that
$t<\inf_{0<m<1} \phi(m) m^{-p}.$
   As in the proof for $p$~even, we can concentrate 
 only on configurations of spins with correlations~$m$
 close to zero, since others bring an exponentially small contribution.
 Note that
$\P(\s\cdot\s'=mN)=\P(\s\cdot\s'=-mN)$ and 
  consequently $I(m)=I(-m)= -m^2/2(1+o(1))$, $m \rightarrow 0$. 
 Summing up the terms in~\eqv(d1) with correlations $m$ and $-m$,
  we  get
$$
\eqalign{   
N^{p-2}\E V_N(t)
=&\frac{2}{\sqrt{2 \pi N}} 
\sum_{m\geq 0, \atop |m| < N^{-1/3-\delta}} N^{p-1}m^p
(e^{t N m^p}-e^{-tN m^p})e^{-N I(m)}- 2 t \E\xi^{2p}+o(1)\cr
 =& \frac{4}{\sqrt{ 2 N \pi}}
\sum_{m\geq 0, \atop |m| < N^{1/3 -\delta} }
 N^{p-1} m^p\big(t N m^p \big)(1+o(1))e^{-N I(m)} - 2t \E\xi^{2p}+o(1)\cr 
   =& \frac{ 4 t }{ \sqrt{ 2  \pi}} 
 \int\limits_{0}^{\infty} s^{2p} e^{-s^2/2} ds - 2
t\E\xi^{2p}+o(1)\rightarrow 0,
\quad  N \uparrow + \infty.\cr 
}
$$    
  Moreover, as for $p$ even, 
 it is also not difficult to show 
   that the truncated value $ N^{(p-2)} \E \widetilde{V}_N(t, \epsilon)$ 
  tends to zero for all $t$ up to  Talagrand's bound~\eqv(c6).       

   Let us now try to perform a rigorous proof of Conjecture~\thv(cj). 
   Proceeding along the lines of the proof for $p$ even, 
  we come to the problem of convergence
   $N^{p-2}\E|V_N(t)| \rightarrow 0.$  
   To get rid of the absolute value of $V_N(t)$, let us first
   apply the Cauchy-Schwartz inequality
  in the same way as it was in  the proof of Proposition~\thv(vwt).
  We obtain
$$
\eqalign{  
\vphantom{\sum}\big[N^{(p-2)} \E|V_N(t)|\big]^{2}
 \leq & \sum\limits_{m_1, m_2, m_3}
(N^{p-1}m_1^p- t\E\xi^p)(N^{p-1}m_2^p-t\E\xi^p) e^{Nt
(m_1^p+m_2^p+m_3^p)}\cr  
&\qquad\qquad{}\times \P(\s\cdot\s'=m_1 N,
 \s\cdot\s'' =m_2 N, \s' \cdot\s ''=m_3 N).
}
\Eq(ssm)  
$$
  Surprisingly, the right-hand side 
  of~\eqv(ssm) does not converge to zero. 
   The problem arises from the fact that
 $$I(m_1, m_2, m_3)= I(-m_1,-m_2, m_3)=I(m_1, -m_2, -m_3)=I(-m_1, m_2,
-m_3),$$ 
  but
  $$ I(m_1, m_2, m_3)\ne  I(m_1, m_2, -m_3).$$  
     In fact,  opening the brackets in 
$(N^{p-1}m_1^{p}-t\E \xi^{2p})(N^{p-1}m_2^{p}-t \E\xi^{2p})$ 
  one can split the right-hand side of \eqv(ssm) 
  into four terms. Let us elaborate the first one
  summing up together the terms with  correlations,
  having the same absolute values $|m_1|, |m_2|, |m_3|$ and 
  the same probability:
$$
\eqalign{
& (2\pi N)^{-3/2}\sum\limits_{m_1>0, m_2>0, m_3>0} N^{(2p-2)}
m_1^{p} m_2^p\, e^{-N I(m_1, m_2, m_3)}\cr
&\qquad\big[e^{t N(m_1^p+ m_2^p+m_3^p)}+e^{t
N(-m_1^p-m_2^p+m_3^p)}
-e^{t N(-m_1^p+ m_2^p-m_3^p)}-e^{t
N(m_1^p-m_2^p-m_3^p)}\big]\cr
 &{}+
 (2\pi N)^{-3/2}\sum\limits_{m_1>0, m_2>0, m_3>0}  N^{(2p-2)} m_1^{p} m_2^p
\,e^{-N I(m_1, m_2, -m_3)}\cr
&\qquad
\big[e^{t N(m_1^p+ m_2^p-m_3^p)}+e^{t N(-m_1^p-m_2^p-m_3^p)}
-e^{t N(-m_1^p+ m_2^p+m_3^p)}-e^{t
N(m_1^p-m_2^p+m_3^p)}\big]\cr
=\,\,& 4 (2 \pi N)^{-3/2}  t \sum_{m_1>0, m_2>0, m_3>0}
         N^{2p-1} m_1^{p} m_2^{p} m_3^{p}(1+o(1)) \, e^{
-N(m_1^2+m_2^2+m_3^2)/2} \cr
&\qquad [e^{-N I(m_1, m_2, m_3)+N
(m_1^2+m_2^2+m_3^2)/2}
 -e^{-N I(m_1, m_2, -m_3)+N
(m_1^2+m_2^2+m_3^2)/2}]\cr
&{}+t^{2}(\E\xi^{2p})^{2}+o(1).
}
$$ 
  This term is of order
$N^{(p-3)/2}t(\E\xi^{p+1})^{3}(1+o(1)) +t^{2} \E \xi^{2p}$,
  since in the expansion of 
$$[e^{-N I(m_1, m_2, m_3)+N (m_1^2+m_2^2+m_3^2)/2}
  -e^{-N I(m_1, m_2, -m_3)+N (m_1^2+m_2^2+m_3^2)/2}],$$ 
  the main term is of order $N m_1 m_2 m_3$.
    The sum of all  other three terms in \eqv(ssm) tends to 
  $ -t^{2}(\E \xi^{2p})^{2}.$ 
 Thus the right-hand side of \eqv(ssm) is of order 
 $N^{(p-3)/2}t(\E\xi^{p+1})^{3}$ and it does not 
 converge to zero for $N\uparrow +\infty$. 
   Therefore the  proof  for $p$ even  is not suitable
  at this point   for $p$ odd.

  A possible solution for  this problem
  is to apply the Cauchy-Schwartz inequality in a different way
  passing to the fourth moment of $Z_N(t)$:    
$$
\eqalign{     
&\big[ N^{(p-2)} \E |V_N(t)| \big]^2
\leq  \E \E_{\s, \s', \s'', \s'''} \Big(N^{p-1}
\Big(R_N(\s,\s')\Big) - t\E\xi^{2p}\Big)
 \Big(\frac{\s'' \cdot\s'''}{N}\Big) - t\E\xi^{2p}\Big)\cr 
&\vphantom{\int}\qquad\qquad\qquad\qquad\qquad\qquad\qquad\qquad{}
\times e^{H_N(t, \s)+
 H_N(t, \s')+ H_N(t, \s'')+H_N(t, \s''') -2Nt}.
}
$$
  It can be proved that the right-hand side of 
   this last inequality tends to zero
  for all $t$ up to some bound.
  But technical details are very tedious.
  We will  only say that
 six parameters
 $m_1, \ldots, m_6$  have to be considered.
  The group of $64$ correlations 
  with fixed absolute values 
   $|m_1|, \ldots, |m_6|$  splits into eight groups
   of correlations having
    the same probabilities. 

 Furthermore, it will be technically even much  harder
  to extend the bound of~$t$  by the truncation
   of the Hamiltonian.
  We will have to take into account five different cases
  and their permutations
  where some of correlations are large and some 
  are small. Each of these cases  will require very tough
  computations.  

\bigskip

\chap{Appendix 2. Two useful theorems}5  

\proposition{\TH(A1)}{\it
 Let $\xi$ be a Gaussian random variable with 
  $\E \xi=0$,  $\E \xi^2=1$. Then for all $a, b>0$
$$
\E[\exp\{a \xi\} \1_{\{\xi>b\}}] \leq \frac{1}{\sqrt{2\pi
(b-a)}}\exp\{-b^2/2+ab\}, \quad \,\hbox{if }b>a,\Eq(stand)
$$
$$
\E[\exp\{a \xi\} \1_{\{\xi<b\}}] \leq \frac{1}{\sqrt{2\pi
(a-b)}}\exp\{-b^2/2+ab\}, \quad \,\hbox{if }b<a.\Eq(stand1)
$$
} 
    
\theo{\TH(thc)}{\it
 Assume that $f(x_1,\ldots,x_d)$ is a function on ${\bf R}^d$ 
 with a Lipschitz constant $L$. Let $J_1,\ldots, J_d$ be 
 independent standard Gaussian random variables. 
Then for any 
 $u>0$ 
$$
\P\{|f(J_1,\ldots, J_d)-\E f(J_1,\ldots, J_d)|>u \}\leq \exp\{-u^2/(2L^2)\}.
\Eq(conc)  
$$        
 }
    
\bigskip\item{[AW]} M.~Aizenman, J.~Wehr, {\it Rounding effects of quenched randomness 
of first-order phase transitions}, {\it Commun. Math. Phys.} {\bf 130}, 489--528, 1990.

\item{[ALR]} M.~Aizenman, J.L.~Lebowitz and D.~Ruelle (1987)
               Some rigorous results on Sherrington-Kirkpatrick
    spin glass model. {\it Commun. Math. Phys.} {\bf 112} 3--20. 
\item{[B1]} A.~Bovier, The Kac version of the Sherrington-Kirkpatrick
            model at high temperatures', 
            {\it J. Stat. Phys.} {\bf 91}, 459-474 (1998)
\item{[B2]} A.~Bovier, Some remarks on the REM and the $p$-spin SK model,
unpublished (1999). 
\item{[BG1]}A. Bovier and V. Gayrard, The retrieval phase of the Hopfield 
            model: A rigorous analysis of the overlap distribution, 
            {\it Probab. Theor. Rel. Fields} 
            {\bf 107}, 61-98 (1996).
\item{[BGP1]}A. Bovier, V. Gayrard and P. Picco, Gibbs states of the
            Hopfield model with extensively many patterns,
            {\it J. Stat. Phys.} {\bf 79}, 395-414 (1995).
\item{[BM]} A.~Bovier and D.~Mason, Extreme value behaviour in the Hopfield 
             model, WIAS-preprint 518,(1999) 
             to appear in {\it Ann. Appl. Probab.} (2001)
\item{[C]} F.~Comets, A spherical bound for the Sherrington-Kirkpatrick model.
 Hommage \`a  P. A. Meyer et J. Neveu.
{\it Ast\'erisque } {\bf   236}, 103-108 (1996).
\item{[CN]} F.~Comets and J.~Neveu (1995) The Sherrington-Kirkpatrick
  Model of Spin Glasses and Stochastic Calculus: The High Temperature
  Case. {\it Commun Math. Phys.} {\bf 166}, 549--564.
\item{[D1]} B.~Derrida, Random energy model: limit of a family of disordered
models, {\it Phys. Rev. Letts.} {\bf 45}, 79-82 (1980).
\item{[D2]} B.~Derrrida, Random energy model: An exactly solvable model of 
disordered systems, {\it Phys. Rev. B} {\bf 24}, 2613-2626 (1981)
\item{[Ei]} TH. Eisele, On a third order phase transition, {\it Commun. 
Math. Phys. }{\bf 90}, 125-159 (1983).
\item{[JS]} J.~Jacod and A.N.~Shiryaev, Limit theorems for stochastic
processes.
      Berlin, Heidelberg, New York: Springer 1987. 
\item{[LLR]} M.~R.~Leadbetter, G.~Lindgren, H.~Rootz\'en,
  {\it Extremes and Related Properties of Random Sequences
    and Processes},
  Springer, Berlin-Heidelberg-New York, 1983.
\item{[NS]}  Ch.M. Newman and D.L. Stein, 
``Thermodynamic chaos and the structure of short range
spin glasses'', in ``Mathematical aspects of spin glasses and neural 
networks'', A. Bovier and P. Picco (Eds.), Progress in Probability,
Birkh\"auser, Boston (1997).
\item{[OP]} E. Olivieri and P. Picco, On the existence of thermodynamics for 
the random energy model, {\it Commun. Math. Phys. }{\bf 96}, 125-144 (1991).
\item{[GMP]} A. Galvez, S. Martinez, and P. Picco, Fluctuations in 
Derrida's random energy and generalized random enery models, {\it J. Stat. 
Phys. }{\bf 54}, 515-529 (1989). 
\item{[RY]} D.~Revuz and M.~Your, Brownian Motion and Continuous 
    Martingales. Berlin, Heidelberg, New York: Springer 1992.
\item{[Ru]} D.~Ruelle,
A mathematical reformulation of Derrida's REM and GREM. 
{\it  Commun. Math. Phys. } {\bf 108}, 225-239 (1987). 
\item{[Shi]} A.N. Shiryaev, A. N. Probability. Second edition.
Graduate Texts in Mathematics, 95. Springer-Verlag, New York, 1996. 
\item{[SK]} D. Sherrington and S. Kirkpatrick, ``Solvable model of a
spin glass'', { \it Phys. Rev. Lett.}
{\bf 35}, 1792-1796 (1972).
\item{[T1]} M.~Talagrand, Rigorous low temperature results 
    for mean field p-spin interaction models. Preprint (1998), to appear in 
Probab. Theor. Rel. Fields.
\item{[T2]} M. Talagrand, ``Concentration of measure and isoperimetric
inequalities in product space'', Publ. Math. I.H.E.S., {\bf 81}, 73-205 
(1995). 

\item{[Tou]} A.~Toubol,  High temperature regime for a multidimensional 
Sherrington-Kirkpatrick model of spin
glass, {\it  Probab. Theor. Rel. Fields }{\bf 110}, 497--534 (1998).
\end